# Nonreciprocal transport in a room-temperature chiral magnet

Daisuke Nakamura[1*#], Mu-Kun Lee[2#], Kosuke Karube[1], Masahito Mochizuki[2], Naoto Nagaosa[1,3], Yoshinori Tokura[1,4], and Yasujiro Taguchi[1,5]

[1]RIKEN Center for Emergent Matter Science (CEMS), Wako 351-0198, Japan.
[2]Department of Applied Physics, Waseda University, Okubo, Shinjuku-ku, Tokyo 169-8555, Japan
[3]Fundamental Quantum Science, TRIP Headquarters, RIKEN, Wako 351-0198, Japan.
[4]Tokyo College and Department of Applied Physics, University of Tokyo, Tokyo 113-8656, Japan.
[5] Baton Zone Program, RIKEN, Wako 351-0198, Japan

Chiral magnets under broken time-reversal symmetry can give rise to rectification of moving electrons, called nonreciprocal transport. Several mechanisms, such as the spin-fluctuation-induced chiral scattering and asymmetry in the electronic band dispersion with and without the relativistic spin-orbit interaction, have been proposed, but clear identification as well as theoretical description of these different contributions are desired for full understanding of nonreciprocal transport phenomena. Here, we investigate a chiral magnet $Co_8Zn_9Mn_3$ and find the nonreciprocal transport phenomena consisting of different contributions with distinct field- and temperature-dependence across the magnetic phase diagram over a wide temperature range including above room-temperature. We successfully separate the nonreciprocal resistivity into different components and identify their mechanisms as spin-fluctuation-induced chiral scattering and band asymmetry in a single material with the help of theoretical calculations.

*daisuke.nakamura.rg@riken.jp
#These two authors contributed equally to this work.

**Introduction**

Chirality in materials encompasses broken space-inversion symmetry of the crystal lattice and provides nonreciprocal responses with respect to the direction of external stimuli when time-reversal symmetry is also broken. For chiral metals, it is known that the voltage drop is different between the opposite current-flow directions (also called the electrical magnetochiral anisotropy effect) [1–4] upon applying a magnetic field. Nonreciprocal electrical transport phenomenon is ubiquitously observed under magnetic fields in various systems with broken space-inversion symmetry, including noncentrosymmetric superconductors [5,6], topological insulator heterostructures [7], and polar semiconductors with Rashba spin-orbit coupling [8]. Such a rectification effect especially in bulk materials may offer a new opportunity to provide a novel electric circuit element.

There are several mechanisms for the nonreciprocal electrical transport phenomena. Helical shape of a macroscopic sample and twisted crystal lattice can induce the self-magnetic-field and chiral scatterings, respectively, under current flow, thereby yielding nonreciprocity [1]. Magnetic materials without inversion symmetry can host chiral spin-ordering, and the nonreciprocal electrical transport has been observed in several chiral magnets with spin-orbit (s-o) interaction [9–12]. In a representative example of MnSi [9], the nonreciprocity in electrical resistivity (hereafter called nonreciprocal resistivity) largely increases just above the Curie temperature ($T_c \sim 29$ K). According to the theoretical explanation based on the Boltzmann equation with s-o coupling, a finite vector spin chirality ($S_i \times S_j$) is produced by the enhanced spin fluctuations around the critical state, which causes the nonreciprocal resistivity [13]. In addition, previous works observed experimentally that the nonreciprocal resistivity appears even below $T_c$ [9–12], and different mechanisms seem to be present, making it difficult to fully understand the nonreciprocal transport phenomena in chiral magnets. Given these previous studies, we focused on a chiral magnet with sufficiently high $T_c$ and observed the nonreciprocal resistivity at room temperature. Moreover, we successfully separated different mechanisms of nonreciprocal resistivity within the single material and theoretically reproduced the characteristic features of the observed nonreciprocal resistivity.

Our target material is a ternary magnet Co-Zn-Mn with the β-Mn-type chiral cubic structure [14–16], whose space group is either $P4_132$ or $P4_332$. The transition temperature $T_c$ to a helical structure of $Co_{10}Zn_{10}$ is around 460 K and gradually decreases with the Mn doping level [15,17]. The $Co_8Zn_9Mn_3$ crystal investigated in this work exhibits $T_c$ close to room temperature. As illustrated in Fig. 1(a), the Wyckoff 8c sites are fully occupied with Co ions and the 12d sites are randomly occupied by Zn and Mn ions [18]. Due to the hyper-Kagome network of magnetic ions in the 12d sites, the magnetic frustration increases with Mn doping level.

For the spin-ordered state below $T_c$, the Dzyaloshinskii-Moriya interaction (DMI) and ferromagnetic exchange interaction compete with each other, resulting in the helical twisting of the spins. By applying a magnetic field, a conical state with a finite magnetization along the field direction appears. In addition, similar to several DMI-induced helimagnets (e.g. MnSi [19]), topologically protected magnetic textures, such as skyrmion [15] and meron/antimeron lattice [20], emerge just below $T_c$. Although the magnetic structure of Co-Zn-Mn has been already established [15–17], electrical transport properties have not been investigated in detail except for magnetoresistance and anomalous Hall resistivity using bulk polycrystals [21,22] and single crystal [23].

**Basic properties**

Figure 1(b) shows the magnetic phase diagram of $Co_8Zn_9Mn_3$, which is obtained from the magnetization curves of a bulk single crystal. The crystal is taken from the same batch of a material used for the FIB-fabricated microdevice (see Supplementary Information Sec. B for details). Below the Curie temperature $T_c$ and critical field $B_c$ into the forced ferromagnetic (FM) state, three phases are identified, namely, helical, conical, and skyrmion phases [Fig. 1(b)]. The critical field $B_c$ and the transition field $B_{H-C}$ from the helical to the conical phases exhibit rapid increases with decreasing temperature below 100 K, which may be influenced by the short-range antiferromagnetic correlation of Mn spins [17]. Because of the cubic symmetry of the crystal lattice, there are three domains with different but equivalent *q*-vectors (*q*//[100], [010], and [001]) in the helical phase, as illustrated in Fig. 1(c). Under sufficiently high magnetic fields above $B_{\text{H-C}}$ along one of the <100> directions, the *q*-vectors align to the magnetic field direction and a single *q*-vector domain of the conical state is realized.

Temperature dependence of the linear resistivity ($\rho_{xx,1f}$; the definition is described in *method*) of the FIB-fabricated microdevice [Fig. 1(d)] is shown in Fig. 1(e). $\rho_{xx,1f}$ gradually decreases as the temperature is reduced, with an anomaly at $T_c = 301$ K, which is evaluated from the $d\rho_{xx,1f}/dT$ curve shown in the inset. The $\rho_{xx,1f}$-$T$ curve is qualitatively similar to that obtained for the bulk crystal but shows somewhat larger value by around 50 %. This may be caused from the residual surface layer of the microdevice inevitably created by the irradiation of Ga-ion beam. As presented in Fig. 1(f), the magnetization (*M*) of bulk crystal at 0.1 T shows typical temperature evolution of ferromagnetic order parameter. Above $T_c$, *M* quickly decreases, indicating a rapid suppression of ferromagnetic or helimagnetic correlations.

**Observation of nonreciprocal electrical transport**

In Figs. 2(a) and 2(b), the magnetic field dependence of the nonreciprocal resistivity ($\rho_{xx,2f}$; the definition is described in *method*) is presented at selected temperatures. Throughout the whole temperature range, $\rho_{xx,2f}$ exhibits clear signals with strong temperature and magnetic field dependence. First, we confirmed that the obtained $\rho_{xx,2f}$ signal indeed represents the nonreciprocal transport by observing a linear relationship between $\rho_{xx,2f}$ value and the current density at 300 K (see Supplementary Information, Sec. D-1), which is expected for the nonlinear nonreciprocal contribution to the resistivity $\rho_{xx,1f}\gamma(B)(\boldsymbol{j}\cdot\boldsymbol{B})/2$. As for the field-angle dependence of $\rho_{xx,2f}$, another component with $\boldsymbol{j}\times\boldsymbol{B}$ dependence is superimposed on the $\boldsymbol{j}\cdot\boldsymbol{B}$ component. However, this component turns out not to be a bulk property, and the results obtained in the $\boldsymbol{j} \parallel \boldsymbol{B}$ geometry presented in the main text are not affected by this term (for details, see Supplementary Information, Sec. D-3).

As shown in Fig. 2(a), for $T < 290$ K, in particular at intermediate temperatures, e.g. 200 K, $\rho_{xx,2f}$ (blue curves) initially shows a positive peak upon increasing the magnetic field from $B = 0$, then decreases and saturates above $B_c$, similarly to the magnetization curves. Therefore, we decompose $\rho_{xx,2f}$ for $T < 290$ K into two components as,

$$\rho_{xx,2f}(T, B) = \rho^{2f}_{scat}(T, B) + \rho^{2f}_{band}(T, B). \qquad (1)$$

The subscripts *scat* and *band* represent scattering-related and electronic-band-asymmetry-related contributions to the nonreciprocal transport, respectively, as discussed later. The first term in Eq. (1) contains the contributions from magnons below $T_c$ and critical fluctuations near $T_c$. The former contribution below 290 K is defined to be proportional to the magnetization at a fixed temperature, $\rho^{2f}_{scat}(T, B) = A_{scat}(T)M(T, B)$, and is determined from the *M-B* curve of bulk crystal, as plotted with the black dashed curves in panel (a). Note that the sign and normalization of $\rho^{2f}_{scat}$ is determined with an assumption that $\rho^{2f}_{band}$, which is depicted as the pink hatched region in panel (a), vanishes in the large magnetic field limit. The residual component in $\rho_{xx,2f}$ for $T < 290$ K is ascribed to $\rho^{2f}_{band}$.

For $T > 290$ K shown in Fig. 2(b), the magnitude of $\rho_{xx,2f}(B)$ initially shows a peak with a negative sign upon increasing the field from $B = 0$, and then its absolute value gradually decreases as the field is further increased. Therefore, we can clearly distinguish the $\rho_{xx,2f}$ signals in this temperature region from the positive contribution $\rho^{2f}_{band}$ observed for $T < 290$ K, and we define them as $\rho^{2f}_{scat}$ (the yellow hatched region). Namely, for $T > 290$ K, we consider

$$\rho_{xx,2f}(T, B) = \rho^{2f}_{scat}(T, B). \qquad (2)$$

To see the overall tendency, we plot the magnitude of $\rho^{2f}_{band}$ in the magnetic phase diagram as a color contour map in Fig. 3(a). $\rho^{2f}_{band}$ becomes larger in the conical phase and gradually decreases as the field is further increased beyond $B_c$. Below 100 K, $\rho^{2f}_{band}$ is suppressed even in the conical phase possibly because of the effect of magnetic disordering due to the evolution of short-range antiferromagnetic correlations of Mn spins [17].

In Fig. 3(b), the color contour map of $\rho^{2f}_{scat}$ is presented in a temperature region close to $T_c$. We note that the color scale in Fig. 3(b) is reversed compared with that in Fig. 3(a), because of the opposite sign between $\rho^{2f}_{scat}$ and $\rho^{2f}_{band}$. The magnitude of $\rho^{2f}_{scat}$ takes maxima just above $T_c$, and shows qualitatively similar $T$ and $B$ dependence to the case of MnSi [9] and the theoretical result based on the scattering from the spin-cluster fluctuations with vector spin chirality [13].

Next, to compare the temperature dependence of $\rho^{2f}_{scat}$ and $\rho^{2f}_{band}$ more quantitatively, we plot the values at $B_{H-C}$ for $\rho^{2f}_{band}$ and the values at 0.1 T (slightly above $B_C$ at $T = 0$ K) for $\rho^{2f}_{scat}$, in Figs. 3(c) and 3(d). Upon decreasing temperature, the magnitude of $\rho^{2f}_{scat}$ starts to evolve at around 330 K, similar to $M(T)$ shown in Fig. 1(f), indicating that the critical spin fluctuation plays a role. It takes a maximum at around $T_c$, then decreases towards low temperatures, and almost vanishes in the limit of zero temperature. On the other hand, $\rho^{2f}_{band}$ starts to increase rapidly below $T_c$, takes a broad maximum around 150 K, decreases gradually, but persists at a finite value (~50 % of its maximum) even in the limit of zero temperature, which is qualitatively different from $\rho^{2f}_{scat}$. These characteristic temperature evolutions in respective contributions suggest different microscopic mechanisms between them.

Here, we classify the nonreciprocal transport from the viewpoints of the s-o interaction and relevance to scatterings. Band asymmetry gives rise to the nonreciprocity, being independent of scatterings. For the spin-ordered states below $T_c$, such as the conical state, the asymmetry in both the band dispersion and magnon scatterings can occur even without the presence of s-o interaction. On the other hand, in the paramagnetic state above $T_c$ or field-induced ferromagnetic state, only when the s-o interaction (namely the DMI) is present, the asymmetric features are allowed from the Boltzmann equation. In our case, the critical spin fluctuations with finite spin chirality just above $T_c$ can induce the nonreciprocal scattering, similar to the case of MnSi [9], and the nonreciprocal resistivity is rapidly reduced upon increasing temperature [Fig. 3(b)], in accord with a sudden drop of the magnetization, as shown in Fig. S2 of Supplementary Information. Below $T_c$, where the long-range ordering is established, magnons are thermally excited as collective spin fluctuations and can give rise to the nonreciprocal scattering. As shown in Fig. 3(d), the $\rho^{2f}_{scat}$ term almost disappears in the zero temperature limit as expected for thermally-induced asymmetric scattering (see also Fig. 4(d) in the latter section). By contrast, as for the intrinsic origin irrelevant to scatterings, the nonreciprocal transport is possible in the conical state even in the zero temperature limit because the asymmetry in the electronic band dispersion persists. The observed features of the $\rho^{2f}_{band}$ term shown in Fig. 3(c), namely, emergence below $T_c$ and persistence at the lowest temperature, are in accord with those expected for the band asymmetry term due to the conical spin structure even without the s-o interaction, and thus identified as such. The detailed theoretical description is presented in the next section and Supplementary Information, Sec. D-4.

For the several contributions of $\rho_{xx,2f}$ with different mechanisms observed in the experiment, we comment on the previous works (as for comparison of the magnitude of nonreciprocal resistivity with different materials, see Supplementary Information Sec.D-2). In MnSi, apart from the strong enhancement of $\rho_{xx,2f}$ in the critical region, $\rho_{xx,2f}$ exhibits a finite value below $B_c$ and saturates far above $B_c$ at the lowest temperature (Supplementary Information in Ref. [9]). Also in $CrNb_3S_6$ [10], $\rho_{xx,2f}$ shows several components, depending on the magnetic field and temperature in the magnetic phase diagram, and the authors assigned them similarly to those based on Eq. (1). They also found another component which is proportional to **B** in the entire temperature region, which is assigned to structural-chirality-induced nonreciprocity. However, in $Co_8Zn_9Mn_3$ and MnSi [9], such a **B**-linear term has not been observed. For the finite nonreciprocity below $T_c$ corresponding to $\rho^{2f}_{band}$ in our work, a possible mechanism is suggested to be the asymmetry in the electronic band dispersion [10], although the theoretical consideration has been missing. Quite recently, the nonreciprocal resistivity originating from asymmetry in the electronic band dispersion is reported in elemental Te [25] and $\alpha$-$EuP_3$ [26]. To clarify the different mechanisms of nonreciprocal transport observed in a single material in the present case, comparison with the corresponding theory is necessary.

**Theoretical analysis**

To explore the various mechanisms of the nonreciprocal transport, we study the Hamiltonian written as

$$H = -t \sum_{i,\mu=x,y,z} c_i^\dagger e^{i\lambda\sigma_\mu/2} c_{i+\mu} + h.c. - \sum_i c_i^\dagger (J\boldsymbol{S}_i - \boldsymbol{B}) \cdot \boldsymbol{\sigma} c_i, \qquad (3)$$

where $c_i^\dagger = (c_{i\uparrow}^\dagger, c_{i\downarrow}^\dagger)$ and $c_i = (c_{i\uparrow}, c_{i\downarrow})^T$ are the electron creation and annihilation operators at position $r_i$, $t$ is the hopping strength, and $\lambda$ is the s-o interaction strength consistent with the DMI in chiral magnets, $\boldsymbol{B} = B\hat{x}$ is the applied field with $B$ being an abbreviation of $g\mu_B B/2$ ($g$ is the g-factor and $\mu_B$ the Bohr magneton), $J(>0)$ is the Kondo coupling strength, $\sigma$ is the Pauli matrix, and $S_i$ is the local magnetizations which in general has the form in conical state as

$$S_i = \sin\alpha\,[\hat{y}\cos(\boldsymbol{q}\cdot\boldsymbol{r}_i) + \eta\hat{z}\sin(\boldsymbol{q}\cdot\boldsymbol{r}_i)] + \cos\alpha\,\hat{x}, \quad (4)$$

where $\boldsymbol{q} = q\hat{x}$ is the helical wavevector, $\eta = \pm 1$ defines the handedness of the helix fixed by the sign of DMI, and $\alpha$ is the tilt angle of the conical state toward the magnetic-field direction $\hat{x}$, as shown in Fig. 4(a).

Equation (3) represents the canonical Hamiltonian for a helical spin structure with the s-o interaction and describes several mechanisms of the nonreciprocal transport which show different magnetic-field and temperature dependence. First, we look at the low-temperature and high-magnetic-field region in Fig. 2(a), where the forced ferromagnetic state is realized with elastic impurity scattering determining the nonreciprocal transport. The asymmetry in the energy dispersion in this case is caused by the s-o interaction and the exchange field. (Note here that the effective Zeeman field is mostly from the exchange field $J$, and scales with the magnetization.) In this limit [$\alpha = 0$ in Eq. (4)], the energy dispersion of Eq. (3) reads

$$\varepsilon_\pm(\boldsymbol{k}) = -2t\cos\frac{\lambda}{2}\sum_{\mu=x,y,z}\cos k_\mu \pm \sqrt{\sum_{\mu=x,y,z}\left[2t\sin\frac{\lambda}{2}\sin k_\mu - Jn_\mu\right]^2}, \quad (5)$$

where $n_\mu$ is the unit vector along the direction of $\boldsymbol{B}$, and we have neglected $\boldsymbol{B}$ compared with $J$. Because there is no obvious contribution in the experimental data which indicates this mechanism, i.e., the band dispersion asymmetry due to the Zeeman field combined with the s-o interaction, we will neglect this in the followings.

The second possible mechanism is the band asymmetry due to the conical spin structure. Putting $\lambda = 0$ and replacing $-2t\sum_{\mu=x,y,z}\cos k_\mu$ by $\frac{k^2}{2m}$, as schematically shown in Fig. 4(b) (also in Supplementary Information, Sec. D-4), this Hamiltonian can be diagonalized with the $k_x$-asymmetric eigenenergy

$$\mathcal{E}_{\pm}^{(\zeta)}(\mathbf{k}) = \frac{k^2 + q^2/4}{2m} \pm \sqrt{\left[\frac{\eta k_x q}{2m} + \zeta(J\cos\alpha - B)\right]^2 + (J\sin\alpha)^2}, \quad (6)$$

where $\zeta = \pm 1$ corresponds to the case with magnetic field applied along the $\pm\hat{x}$ direction, while we keep $B$ and $\cos\alpha$ positive in both cases.

By using the Boltzmann transport theory with the single-relaxation-time approximation [8,27] (see Supplementary Information, Sec. D-4), we obtain the second-order nonreciprocal resistivity which we identify with $\rho^{2f}{}_{band}(B)$ in the lowest order of $q$ at zero temperature as

$$\rho^{2f}{}_{band}(B) = \frac{27\pi^4 \zeta \eta j_0 q^3 (k_{F,+}^3 - k_{F,-}^3)}{4\tau e^3 J^2 (k_{F,+}^3 + k_{F,-}^3)^3} \cos\alpha(B) \sin^2\alpha(B), \quad (7)$$

where $k_{F,\pm} = \sqrt{2m(\mu \mp J)}$ up to zeroth order of $B$ and $q$, $\mu$ is the chemical potential, $\tau$ is the relaxation time, and $e$ is the absolute electron charge.

In Fig. 4(c), we plot the calculated $\rho^{2f}{}_{band}$ as a function of $B$. The qualitative behavior is reproduced, i.e., it has the maximum. Note that for $B > 0.1$ T, theoretical $\rho^{2f}{}_{band}$ becomes zero because we assumed $\alpha(B = 0.1\text{ T}) = 0$. There is remaining finite $\rho^{2f}{}_{band}$ observed in our experiment in this nearly ferromagnetic regime especially at higher temperatures, which may be the consequence of the empirical assumption for the field dependence of $\rho^{2f}{}_{scat}(T, B) = A_{scat}(T)M(T, B)$ below 290 K.

For $\rho^{2f}{}_{scat}(T, B)$, we can consider the third mechanism of the nonreciprocal transport, i.e., asymmetric scattering due to the spin fluctuation with vector spin chirality, which we refer to as $\rho^{2f}{}_{scat}$. Basically, the spin fluctuation at low temperatures can be described by the magnon theory in the forced ferromagnetic state, which we describe in the Supplementary Information D-4.4. In Fig. 4(d), we show the temperature dependence of the vector spin chirality due to the DMI produced by thermally excited magnons. The thermal distribution of the magnons is also plotted for comparison in Fig. S13. As for $\rho^{2f}{}_{scat}$ above 290K, the magnon is not appropriate to describe the spin fluctuation and the mode coupling approximation has been developed to describe the situation [13]. The resulting $T$ and $B$ dependence are very similar to those for $\rho^{2f}_{scat}$, which support our identification.

**Conclusion and outlook**

We investigated nonreciprocal electrical transport phenomena in a room-temperature chiral magnet $Co_8Zn_9Mn_3$ for the *I*//*B* configuration. Two contributions of the nonreciprocal resistivity are found in different temperature and magnetic-field regimes. The first one, which dominantly appears in the forced ferromagnetic state or paramagnetic state in the critical regime, is identified as being due to the nonreciprocal scattering induced by magnons or spin-chirality critical fluctuations. The second one, which is most pronounced in the conical state, is attributed to the electronic band asymmetry due to the exchange coupling between conduction electrons and the conical-spin background. The observed field and temperature dependence is qualitatively in accord with the prediction of the Boltzmann theory. The separation of nonreciprocal resistivity into two components with different $T$ and $B$ dependences in a single material enabled to identify their microscopic mechanisms. Our experimental observation and subsequent classification of different contributions based on theoretical consideration provide a great step towards comprehensive understanding of the nonreciprocal transport phenomena in chiral magnets.

**Methods**

Single crystal of $Co_8Zn_9Mn_3$ was grown by the Bridgman method. The chemical composition of the single crystal was evaluated to be $Co_{7.72}Zn_{9.40}Mn_{2.88}$ by scanning electron microscope-based energy dispersive X-ray spectroscopy. Because the deviation from the nominal value is small, we refer to this crystal as $Co_8Zn_9Mn_3$ for simplicity throughout the paper. The micro-sample with 14 μm × 4 μm in lateral size and 0.5 μm in thickness was fabricated and connected to the six-terminal lead pads by a focused ion beam (FIB) instrument (Hitachi, NB-5000), as shown in Fig. 1(d).

The electrical transport properties were measured by using a source meter (Keithley, 6221) and lock-in-amplifiers (Standard Research, SR865 and NF corporation, LI5650). The magnetic field and temperature were controlled by Physical Properties Measurement

System (Quantum Design). The nonreciprocal resistivity is described as the second term in the measured electrical resistivity, $\rho = \rho_{xx,1f} + \rho_{xx,1f}\gamma(B)(\boldsymbol{j}\cdot\boldsymbol{B})/2$, where $\rho_{xx,1f}$ is the linear resistivity and $\gamma$ is the coefficient of nonreciprocal transport. When the AC current density $\boldsymbol{j} = \boldsymbol{j}_0\sin\omega t$ is applied, the output voltage from the second term of the resistivity, $\rho_{xx,1f}\gamma(B)(\boldsymbol{j}\cdot\boldsymbol{B})\boldsymbol{j}/2$, is proportional to $\sin(\pi/2 - 2\omega t)$ (plus a constant). Therefore, we define the nonreciprocal resistivity ($\rho_{xx,2f} = \rho_{xx,1f}\gamma(B)(\boldsymbol{j}\cdot\boldsymbol{B})/2$) as the imaginary part of the second harmonic resistivity. The typical values of AC current density and frequency for $\rho_{xx,2f}$ measurements were $3\times10^9$ A/m$^2$ and 377 Hz, respectively. Then, the output signal is antisymmetrized with respect to the magnetic field to remove the other contributions. Details are described also in Supplementary Information, Sec. A.

**Acknowledgements**

We thank Y. Onose for fruitful discussions. D.N. thank Y.-L. Chiew, K. Nakajima, X. Z. Yu and RIKEN CEMS Emergent Matter Science Research Support Team for technical assistance of preparing microdevices. This work was financially supported by JST CREST (Grant No. JPMJCR20T1 and JPMJCR1874), JSPS Grant-in-Aids for Scientific Research (Grant no. 20H00337, 23K26534, 24H00197 and 24H02231), Waseda University Grant for Special Research (Grant No. 2024C-153), the RIKEN TRIP initiative (Many-body Electron Systems, Advanced General Intelligence for Science Program, and Fundamental Quantum Science Program), and Sumitomo Chemical.

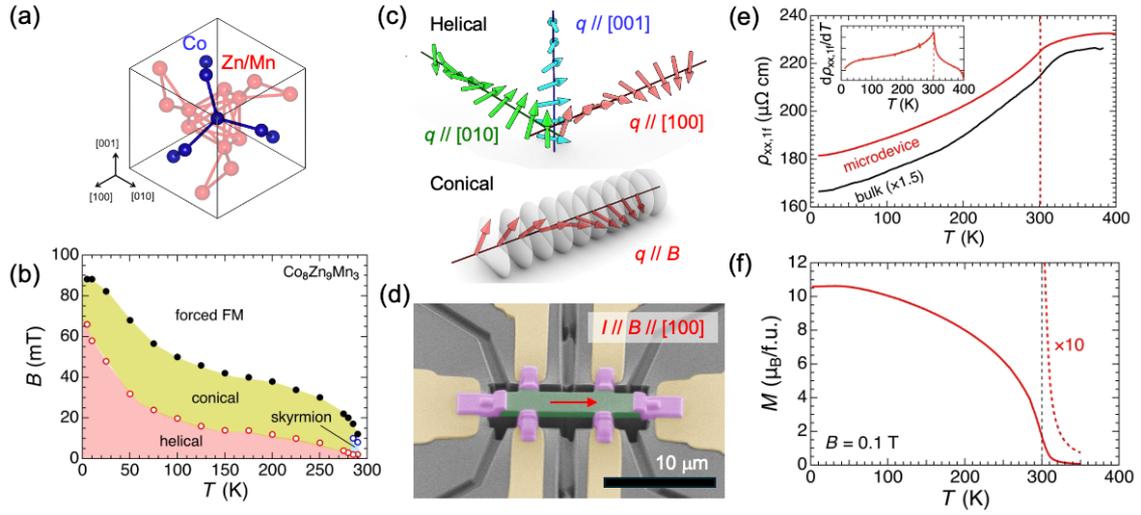

**Fig. 1: Characteristics of Co$_8$Zn$_9$Mn$_3$.**

(a) β-Mn type crystal structure of Co-Zn-Mn chiral magnet. The atoms at 8c and 12d sites are colored with blue and red, respectively. (b) Magnetic phase diagram of Co$_8$Zn$_9$Mn$_3$, obtained from the magnetization curves for a bulk crystal with the same aspect ratio as the microfabricated sample. (c) Schematic spin arrangement in helical and conical phases. (d) Scanning electron microscopy (SEM) image of the FIB-fabricated microdevice. The scale bar represents 10 μm. The image is colored to distinguish Co$_8$Zn$_9$Mn$_3$ (green), tungsten electrodes (pink) and gold pads (yellow). (e) Temperature dependence of the linear resistivity ($\rho_{xx,1f}$) of the FIB-fabricated microdevice (red) and bulk crystal (black). The vertical dotted line indicates $T_c$ = 301 K, which is evaluated from the anomaly in the d$\rho_{xx,1f}$/d$T$ curve shown in the inset. (f) Temperature dependence of magnetization ($M$) of the bulk crystal measured in a warming run after the field-cooling condition, under application of a magnetic field $B$ of 0.1 T. The red dashed curve for $T > T_c$ shows the enlarged view of the $M(T)$ in the critical regime.

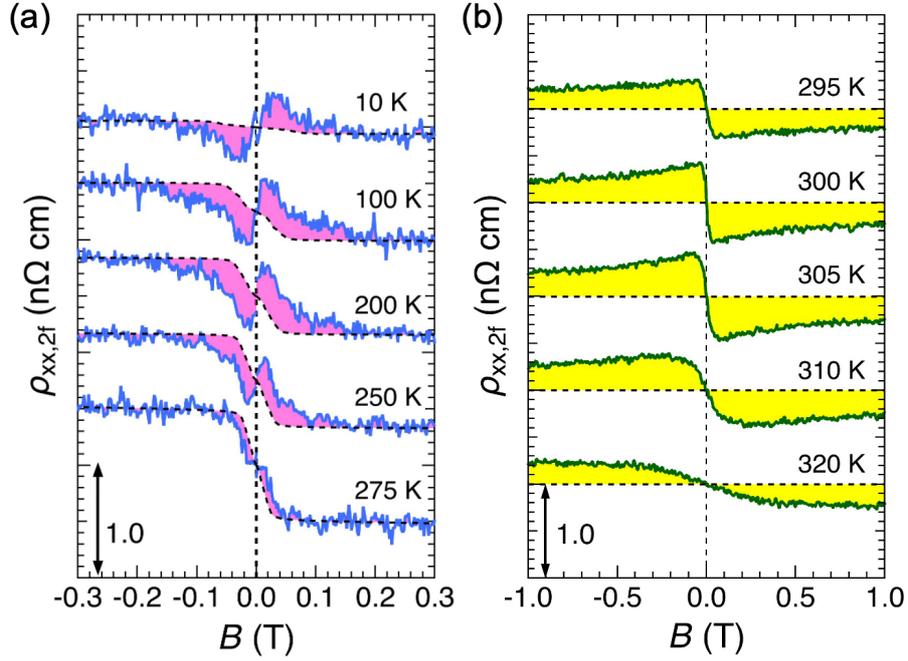

**Fig. 2: Nonreciprocal transport observed in $Co_8Zn_9Mn_3$.**

(a), (b) Magnetic field dependence of nonreciprocal resistivity ($\rho_{xx,2f}$) at selected temperatures below 275 K [(a), blue curves] and above 295 K [(b), green curves]. The black dashed curves in (a) represent the component (magnon-induced $\rho^{2f}_{scat}$) proportional to ***M***, which is estimated from the ***M-B*** curve of bulk crystal with the same aspect ratio with the microfabricated sample. The pink- and yellow-hatched regions correspond to the magnitude of the different contributions $\rho^{2f}_{band}$ and chiral-fluctuation-induced $\rho^{2f}_{scat}$, respectively, in the nonreciprocal resistivity. The values of AC current density and frequency for $\rho_{xx,2f}$ measurements were $3\times10^9$ A/m$^2$ and 377 Hz, respectively.

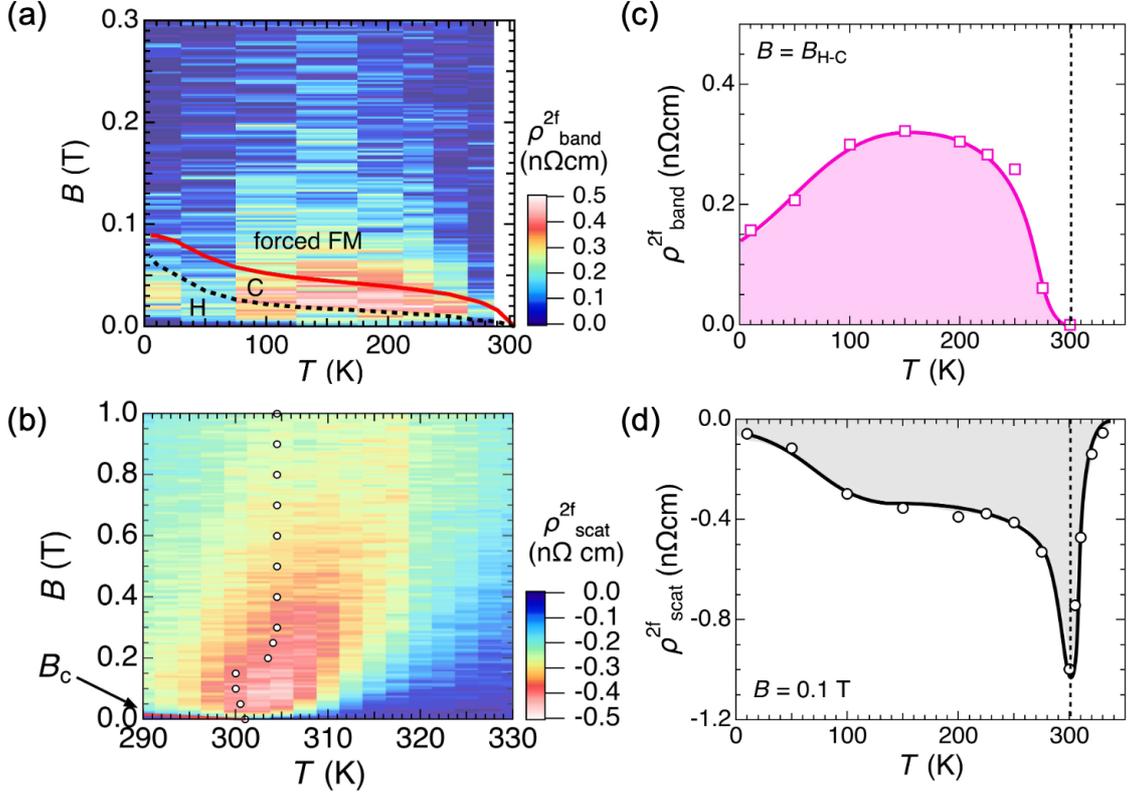

**Fig. 3: Temperature profiles of decomposed contributions to the nonreciprocal resistivity.**

(a), (b) The color contour maps of the nonreciprocal resistivity contributions $\rho^{2f}_{band}$ and $\rho^{2f}_{scat}$ in the magnetic phase diagram. In (a), H and C stand for the helical and conical spin states, respectively. In (b), the color scale is reversed to that of (a) due to the opposite sign between $\rho^{2f}_{band}$ and $\rho^{2f}_{scat}$. The open circles indicate ferromagnetic-to-paramagnetic crossover fields evaluated from the peak positions of $d\rho_{xx,1f}/dT$ curves. (c), (d) Temperature profiles of the two contributions of the nonreciprocal resistivity, $\rho^{2f}_{band}$ (c) and $\rho^{2f}_{scat}$ (d). The vertical dotted line indicates $T_c$.

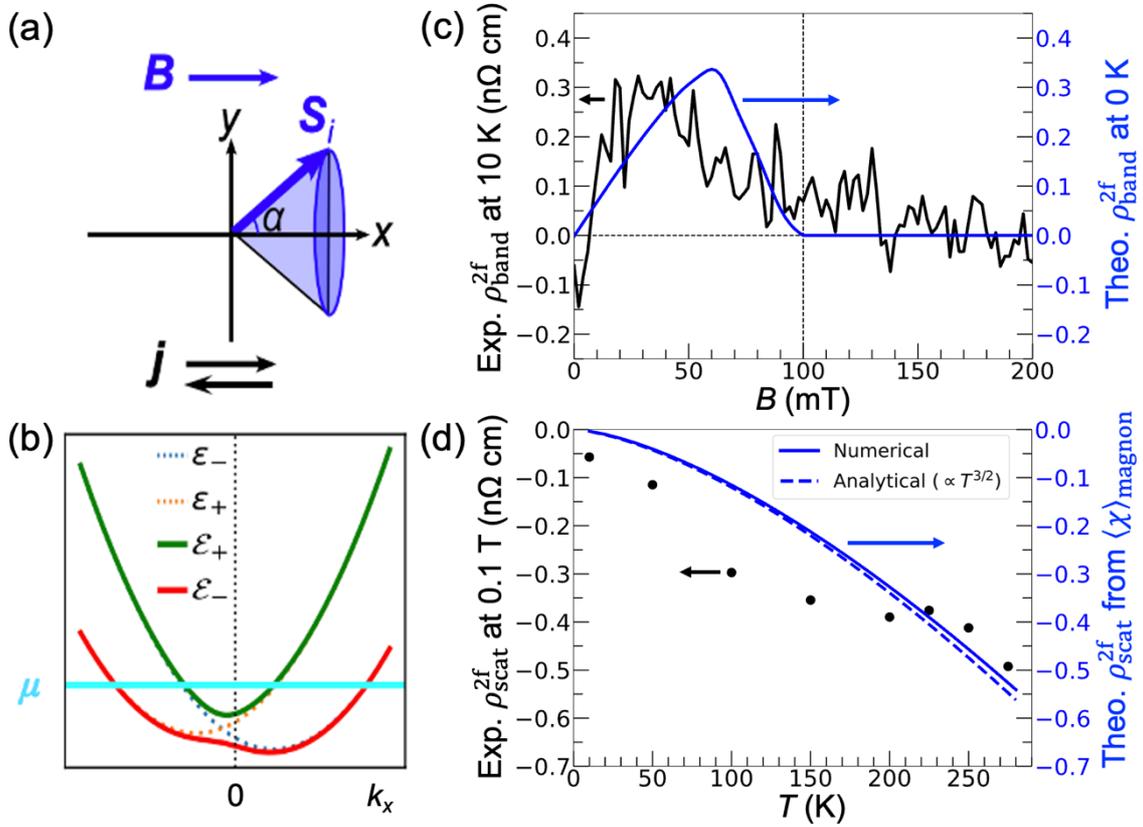

**Fig. 4: Theoretical results of the nonreciprocal resistivity.**

Schematics of (a) the conical spin state and (b) the asymmetric electronic band dispersion relations in a magnetic field applied in the $+\hat{x}$ direction. (c) Comparison of experimental data of $\rho^{2f}_{band}$ observed at 10 K (black curve) and the trend predicted by our theory at 0 K (blue curve), which is the function $\cos\alpha(B)\sin^2\alpha(B)$ using values of $\alpha(B)$ extracted from Fig. S12 in Supplementary Information. (d) Comparison of experimental data of $\rho^{2f}_{scat}$ observed at 0.1 T [black dots from Fig. 3(d)] for $T < 280$ K and theoretical results (blue dotted and solid curves) calculated by the magnon-induced vector spin chirality, with the approximate analytical value being proportional to $T^{3/2}$ (see Supplementary Information D-4.4 for details).

**Data availability**

Datasets generated in the present work are available from the corresponding authors on reasonable request.

**Contributions**

D.N., Y.Tokura and Y.Taguchi coordinated the project. K.K. grew the single crystal of Co$_8$Zn$_9$Mn$_3$. D.N. prepared the microfabricated device and performed the nonreciprocal transport measurements. D.N. and K.K. performed the magnetization measurements. D. N. analyzed the experimental data. M.-K.L., M.M. and N.N. developed the theoretical framework on nonreciprocal resistivity. D.N., M.-K.L., N.N. and Y. Taguchi wrote the manuscript with the inputs from all co-authors.

**Corresponding author**

Correspondence to Daisuke Nakamura.

# Supplementary Information of
## "Nonreciprocal transport in a room-temperature chiral magnet"

**Contents**

**A: Experimental setup**

**B: Magnetization data and phase diagram**

**C: Linear resistivity**

**D-1: Nonreciprocal resistivity – Bias dependence**

**D-2: Nonreciprocal resistivity – Nonreciprocal coefficient**

**D-3: Nonreciprocal resistivity – Field-angle dependence**

**D-4: Nonreciprocal resistivity – Theoretical calculation of second harmonic resistivity**

## A: Experimental setup

The chemical composition of the bulk crystal from which microdevices are fabricated by focused ion beam (FIB) is evaluated by scanning electron microscope-based energy dispersive X-ray spectroscopy (SEM-EDX). The several positions around the milled area for the FIB microdevice are investigated as shown in Fig. S1(a), and the atomic ratio of Co, Zn and Mn at each position is plotted in Fig. S1(b). The dotted lines indicate nominal values for $Co_8Zn_9Mn_3$. After subtracting the extrinsic effect from the injected Ga ion of FIB process, the chemical composition of bulk crystal is determined to be $Co_{7.72}Zn_{9.40}Mn_{2.88}$.

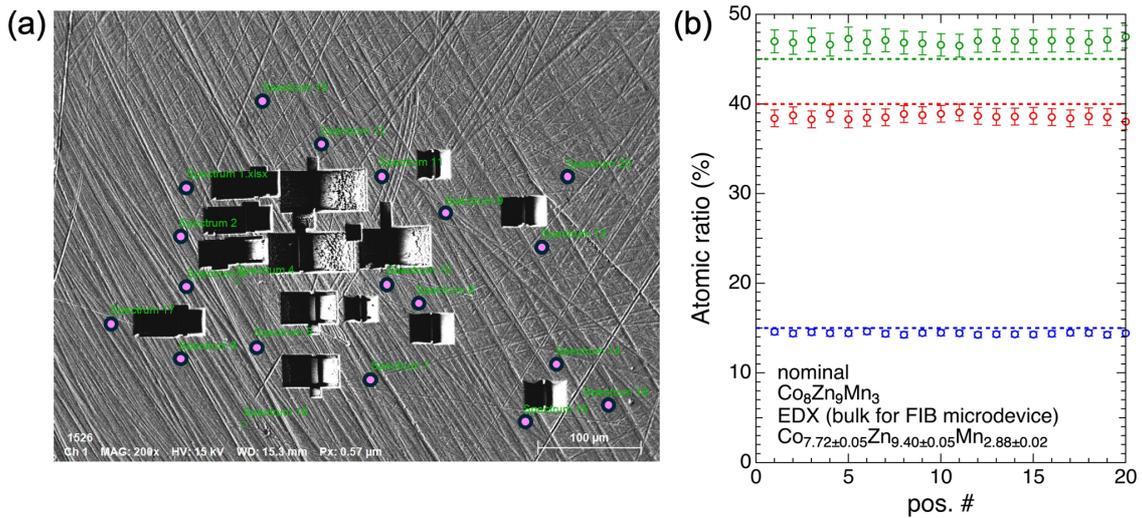

**Fig. S1:** (a) SEM image of the bulk $Co_8Zn_9Mn_3$ crystal for fabricating FIB microdevice. The pink circles indicate the positions where the EDX spectra were measured. (b) The position-dependent atomic ratio evaluated from SEM-EDX.

After fabricating the microdevice by the FIB instrument, $Al_2O_3$ layer with 5 nm thickness was deposited at room temperature by the atomic layer deposition instrument (Fiji, F200), to prevent the surface reaction to the air. Upon measuring the nonreciprocal resistivity at different temperatures, the sample temperature is once increased above $T_c$ under zero magnetic field, and then set to each measurement temperature, in order to minimize the possible history effect. The raw data of $\rho_{xx,2f}$ are antisymmetrized against the magnetic field, $\rho_{xx,2f}(B)/2 - \rho_{xx,2f}(-B)/2$, to exclude the extrinsic background. The size and $T_c$ of the samples used in this work, and the typical current density are listed in Table S1. The data in the main text is obtained for sample #1.

|  | $X$ (μm) | $Y$ (μm) | $t$ (μm) | $x$ (μm) | $T_c$ (K) | $j$ (GA/m$^2$) |
|---|---|---|---|---|---|---|
| $Co_8Zn_9Mn_3$ - #1 | 14.2 | 3.50 | 0.42 | 5.15 | 301 | 2.0 |
| $Co_8Zn_9Mn_3$ - #2 | 14.2 | 2.32 | 0.43 | 5.55 | 300 | 2.1 |
| $Co_8Zn_9Mn_3$ - bulk | 3710 | 910 | 110 | - | 296 | - |

**Table. S1:** List of sample size, $T_c$ and typical current density $j$ for $\rho_{2f}$ measurements. $X$ and $Y$ are the lateral size of thin plate, $t$ is the thickness and $x$ is the distance between the voltage electrodes.

**B: Magnetization data and phase diagram**

The magnetization of $Co_8Zn_9Mn_3$ is investigated to compare the field-evolution of the nonreciprocal resistivity and the magnetic phase diagram. Because the size of the microdevice used in the nonreciprocal resistivity measurements is too small to detect a sufficient magnitude of the magnetization by a SQUID magnetometer, we prepared a bulk piece of single crystal from the same batch as used for the microdevices. In order to minimize the difference in the demagnetization effect between bulk and microdevice, the bulk crystal is polished to the thin plate with almost the same aspect ratio as shown in Table S1. The magnetization is measured by using MPMS (Quantum Design).

The temperature dependence of the magnetization under the magnetic field of 20 Oe is presented in Fig. S2. The overall temperature dependence is similar to that reported previously [16]. From the minimum in the temperature derivative curves shown in the inset, the Curie temperature ($T_c$) is evaluated to be 296 K. As compared in Table S1, $T_c$s of bulk crystal and the microdevices, the latter of which are determined by resistivity measurements, are almost the same.

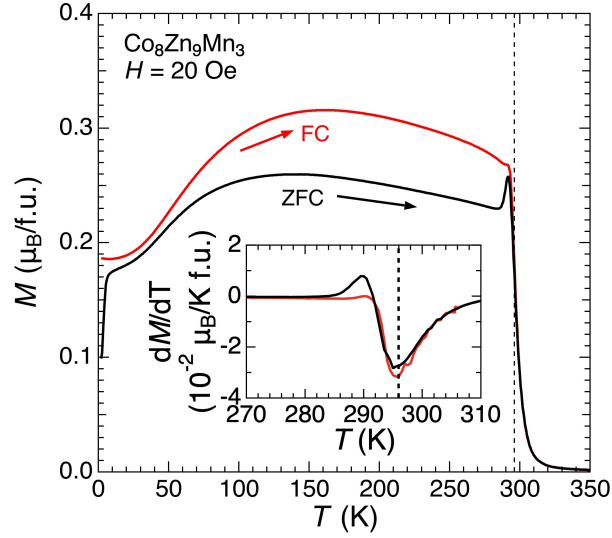

**Fig. S2:** Temperature dependence of the magnetization in $Co_8Zn_9Mn_3$ measured in warming runs after the field-cooling (FC) and zero-field-cooling (ZFC) conditions. The inset represents the enlarged plot of $dM/dT$ curve. The dotted lines indicate the critical temperature, which is defined as the minimum of $dM/dT$.

The magnetic field dependence of the magnetization is shown in Fig. S3 at (a) 285 K, (b) 100 K and (c) 25 K. The bottom, middle and top panels present $M$, the first derivative and the second derivative of $M$ against the magnetic field, respectively, in the field-up-sweep process. The phase transition fields are determined as the extremal points in $d^2M/dB^2$ curves, and the magnetic phase diagram is presented in Fig. S3(d). The helical (hatched in pink), conical (dark yellow), skyrmion (pale blue) and forced ferromagnetic states (white) are identified. The transition fields into forced ferromagnetic state ($B_c$), and from helical to conical states ($B_{H-C}$), start to rapidly increases below 100 K. This suggests that the magnetic disorder appears at low temperatures, due to the development of antiferromagnetic short-range correlation of Mn spins. For the higher Mn doping levels, the spin glass state appears more prominently at low temperatures typically around 60 K [18].

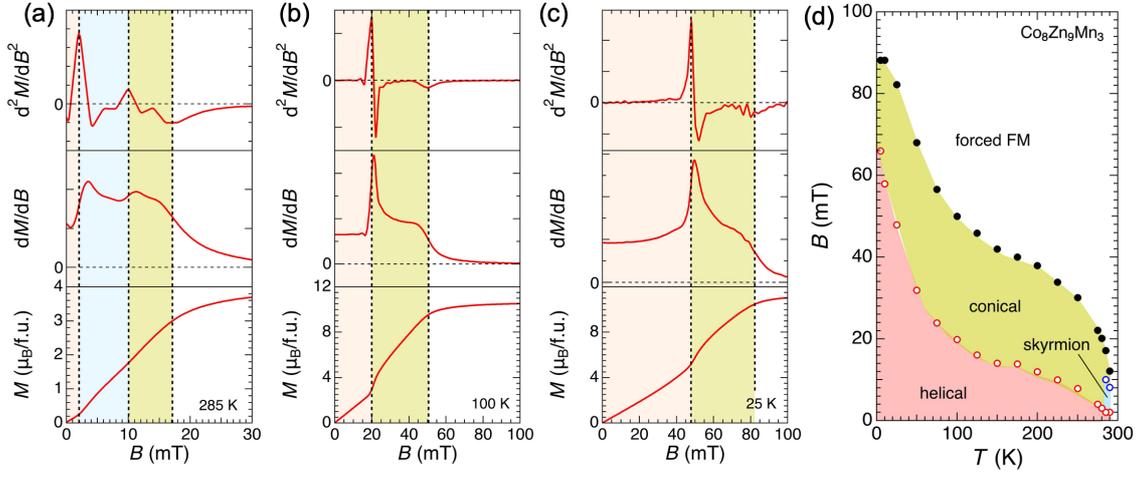

**Fig. S3:** (a)–(c) Magnetic field dependence of the (bottom panel) magnetization, (middle panel) $dM/dB$ and (top panel) $d^2M/dB^2$ at (a) 285 K, (b) 100 K and (c) 25 K, respectively. The boundaries between different phases presented by different color regions are determined as the peaks of $d^2M/dB^2$ curve. (d) Magnetic phase diagram.

**C: Linear resistivity**

In this section, the linear longitudinal and Hall resistivity of the microdevice used in this work are presented. The magnetic field dependence of the Hall resistivity $\rho_{yx}$ with the $I\perp B$ configuration shown in Fig. S4(a) exhibits a gradual increase above $T_c$ and a saturation at low temperatures, indicating that the anomalous Hall effect is dominant. The normalized linear magnetoresistance by its value at zero magnetic field ($\rho_{xx}/\rho_{xx, B=0}$) is plotted in Figs. S4(b) and (c), for the $I\perp B$ and $I//B$ configurations, respectively. In (b), $\rho_{xx}/\rho_{xx, B=0}$ slightly decreases with the magnetic field, indicating the suppression of the electron scattering from the fluctuating spins. We note that $\rho_{xx}/\rho_{xx, B=0}$ becomes largest at 300 K (~$T_c$), due to enhanced spin fluctuations in the critical region. For $I//B$ configuration, there are kink structures near $B_{H-C}$ (black arrows) determined in Fig. S3.

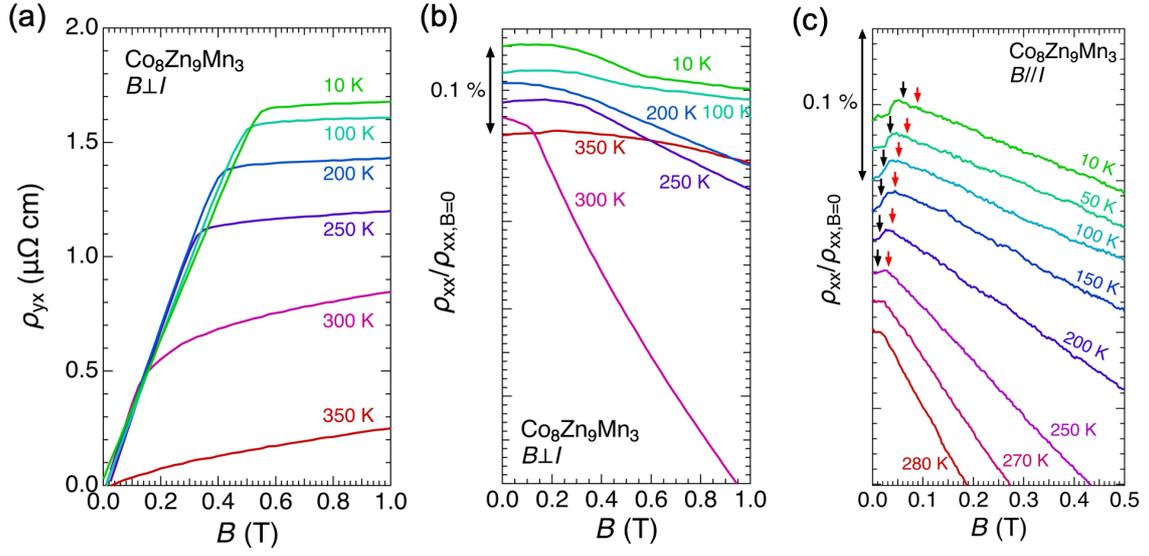

**Fig. S4:** (a) Magnetic field dependence of Hall resistivity at selected temperatures, with the $I \perp B$ configuration. (b) Magnetic field dependence of linear resistivity, with the $I \perp B$ configuration. (c) Magnetic field dependence of linear resistivity, with the $I // B$ configuration. The black and red arrows indicate $B_{H-C}$ and $B_c$ evaluated in Fig. S3.

**D-1: Nonreciprocal resistivity – Bias dependence**

Because the nonreciprocal resistivity is a nonlinear response to the electrical current, $\rho_{xx,2f}$ is proportional to the injected current density, $j$. As shown in Fig. S5(a), we measured the $j$ dependence of $\rho_{xx,2f}$ at 300 K, where $\rho^{2f}_{scat}$ (due to spin chirality fluctuation $S_i \times S_j$) component is observed in Fig. 3(b) of the main text. The peak structure around 0.05 T and the value at 1.0 T increase with $j$, indicating that $\rho^{2f}_{scat}$ is a nonlinear phenomenon. In the bottom panel of Fig. S5(b), we confirm the $j$-linear increase of the maximum value of $\rho_{xx,2f}$ that represents the spin fluctuation component $\rho^{2f}_{scat}$. The linear resistivity ($\rho_{xx,1f}$) also increases very slightly with changing $j$ due to the Joule heating effect, as plotted in the top panel of Fig. S5(b). From the increment of $\rho_{xx,1f}$ and the $\rho_{xx,1f}(T)$ curve shown in Fig. 1(e) of the main text, the temperature increase of the microdevice at $j = 3.0 \times 10^9$ A/m$^2$ is evaluated to be ~ 3 K.

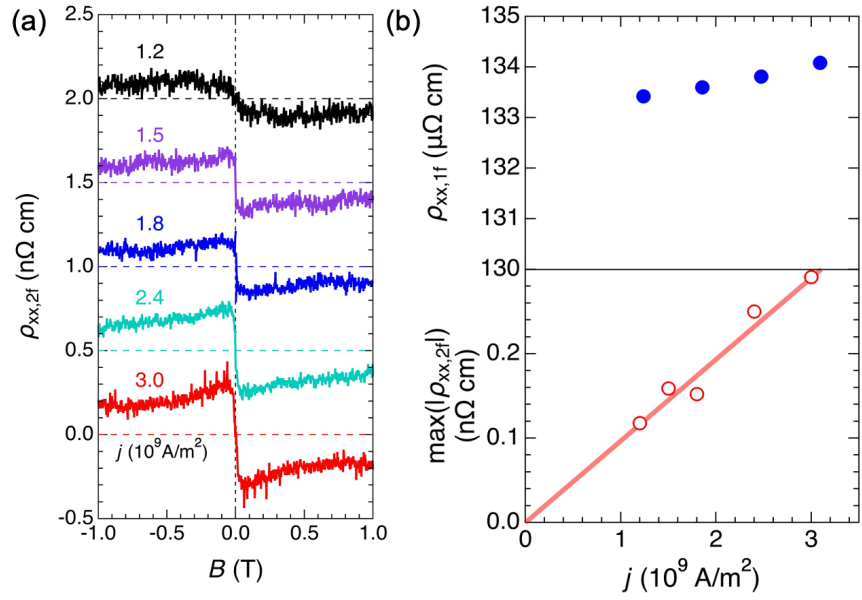

**Fig. S5:** (a) Magnetic field dependence of nonreciprocal resistivity at selected electrical current densities, measured at 300 K. The vertical offset is added to make a clearer view. (b) Electrical current density dependence of (bottom) the maximum value of the nonreciprocal resistivity evaluated from (a) and (top) linear resistivity at 300 K. In the bottom panel, the bold line is a linear fitting result.

**D-2: Nonreciprocal resistivity − Nonreciprocal coefficient**

To compare the observed magnitude of the nonreciprocal resistivity with different materials, we focus on the nonreciprocal coefficient γ, evaluated from $\rho_{xx,2f} = \rho_{xx,1f}\gamma(B)(\boldsymbol{j}\cdot\boldsymbol{B})/2$. For calculation, we neglect the magnetic field dependence of the linear longitudinal resistivity $\rho_{xx,1f}$, due to its small magnitude (~ 0.1 %, Fig. S4(c)), as compared with the magnitude of temperature evolution (~ 20 %, Fig. 1(e)). In Fig. S6(a), the color contour plot of the absolute value of |γ| is presented, because $\rho_{xx,2f}$ changes its sign depending on temperature. We find strong enhancement of |γ| in the critical region around 300 K, which is contributed from $\rho^{2f}_{scat}$. From the line cut of the color contour plot at 302.5 K, we evaluate the maximum value of |γ| in $Co_8Zn_9Mn_3$ to be $1.6 \times 10^{-13}$ $m^2$/TA at $B$ = 3 mT. This value is smaller than other helimagnets; e.g., $3\times 10^{-12}$ $m^2$/TA for MnSi [9] and $\sim 10^{-12}$ $m^2$/TA for $CrNb_3S_6$ [10]. This is probably due to the large helical periodicity in $Co_8Zn_9Mn_3$ ($\lambda$ = 102 nm around $T_c$ [S1]) compared with $\lambda \sim$ 18 nm in MnSi [24], which may cause a smaller magnitude of vector spin chirality and results in smaller γ.

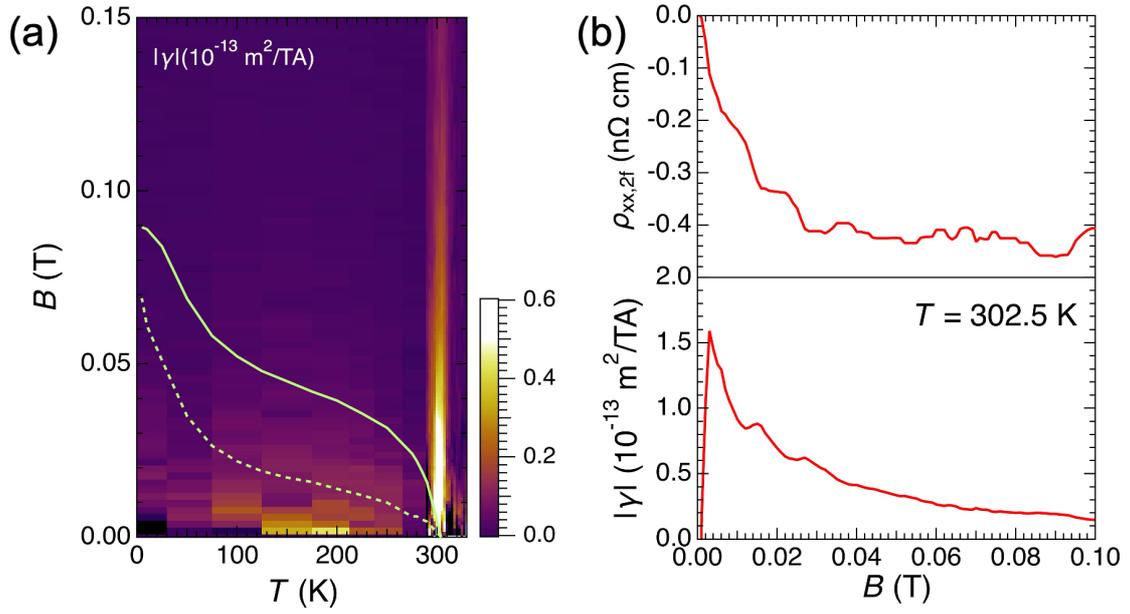

**Fig. S6:** (a) Color contour plot of the absolute value of nonreciprocal coefficient, $|\gamma|$. The solid and dashed curves are the phase transition field $B_c$ and $B_{H\text{-}C}$, determined from the magnetization curves shown in Fig. S3. (b) The line cut of (top panel) $\rho_{xx,2f}$ and (bottom panel) $|\gamma|$ at $T = 302.5$ K.

**D-3: Nonreciprocal resistivity − Field-angle dependence**

For chiral magnets, the relationship $\rho_{xx,2f} \propto \boldsymbol{I} \cdot \boldsymbol{B}$ is expected when the time reversal symmetry is broken by the external magnetic field, and hence $\rho_{xx,2f}$ is supposed to take a maximum value in the $\boldsymbol{I}//\boldsymbol{B}$ configuration. Therefore, the magnetic field angle dependence of $\rho_{xx,2f}$ is investigated in this section. As presented in the following, an additional term showing $\boldsymbol{I} \times \boldsymbol{B}$ dependence is observed to coexist, but this component turns out not to be an intrinsic property of the bulk, and possibly arises from the interface with the electrode.

For a series of measurements, another microdevice #2 (Fig. S7(a)) was fabricated from the same bulk single crystal as for the microdevice #1 shown in the main text. As shown in Fig. S7(b), the linear resistivity of the microdevice #2 exhibits slightly lower than that of #1 [Fig. 1(e) of the main text]. We rotate the magnetic field direction in the *xy*-plane (i.e. while keeping $\theta = 90$ deg.) of Fig. S7(c) and the azimuth angle ($\phi$) dependence is measured. $\rho_{xx,2f}(\phi) = [\rho_{xx,2f}(B,\phi) - \rho_{xx,2f}(-B,\phi)]/2$ is obtained at selected temperatures and magnetic fields in the helical, conical and forced ferromagnetic phases, as indicated by the red circles in Fig. S7(d).

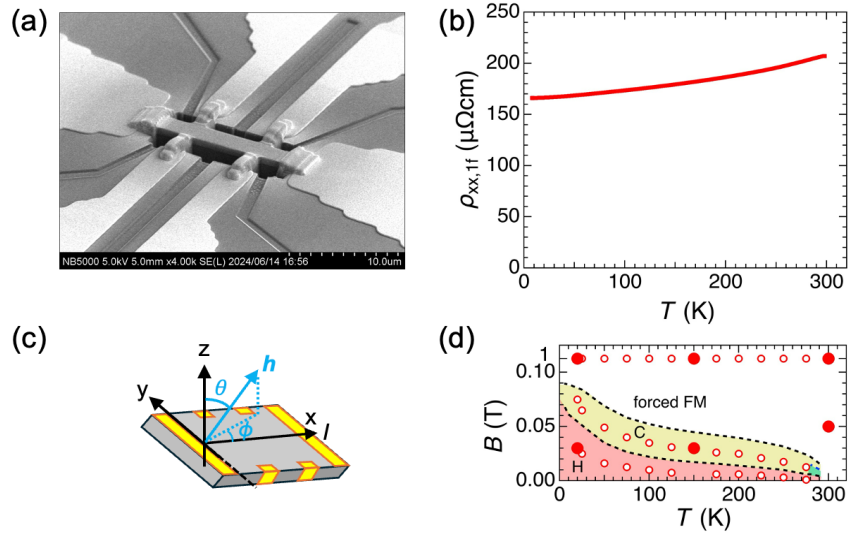

**Fig. S7:** (a) SEM image of $Co_8Zn_9Mn_3$ microdevice #2, used for the field-angle dependence of $\rho_{xx,2f}$ measurements. (b) Temperature dependence of linear longitudinal resistivity in microdevice #2. (c) Schematic illustration of the field-angle dependent measurement of nonreciprocal resistivity. (d) Magnetic fields and temperatures for the field-angle dependent measurements are indicated as red circles in the magnetic phase diagram [the same as Fig. 1(b) in the main text]. The results at filled and open circles are plotted in Figs. S8 and S9, respectively.

The $\rho_{xx,2f}(\phi)$ data at the representative temperatures and magnetic fields [solid circles in Fig. S7(d)] are shown in Fig. S8. The open symbols in the bottom panels are the experimental data, and the bold gray curves are the fitting function $A_1\sin\phi + A_2\cos\phi$ that represents the summation of $I \times B$ and $I \cdot B$ components. The nonreciprocal resistivity presented in the main text is measured at $\phi = 0$ deg., indicating that only $I \cdot B$ component is detected. On the other hand, $\rho_{xx,2f}(\phi)$ in Fig. S8 does not take a maximum value at $\phi = 0$, suggesting that $I \times B$ component also exists in $\rho_{xx,2f}(\phi)$, in addition to the $I \cdot B$ components ($\rho^{2f}_{scat}$ and $\rho^{2f}_{band}$) described in the main text. The decomposed $I \cdot B$ and $I \times B$ components in the fitting function are depicted in the top panels as red and blue curves, respectively. The ratio between the amplitudes of $I \cdot B$ and $I \times B$ components is also described. The $I \times B$ component becomes dominant in $\rho_{xx,2f}(\phi)$ at 1 T far above $B_c$ [Figs. S8(b), S8(d) and S8(f)], whereas it is significantly suppressed in the low field region in the spin fluctuation regime [Fig. S8(a)], the conical [Fig. S8(c)] and helical [Fig. S8(e)] states.

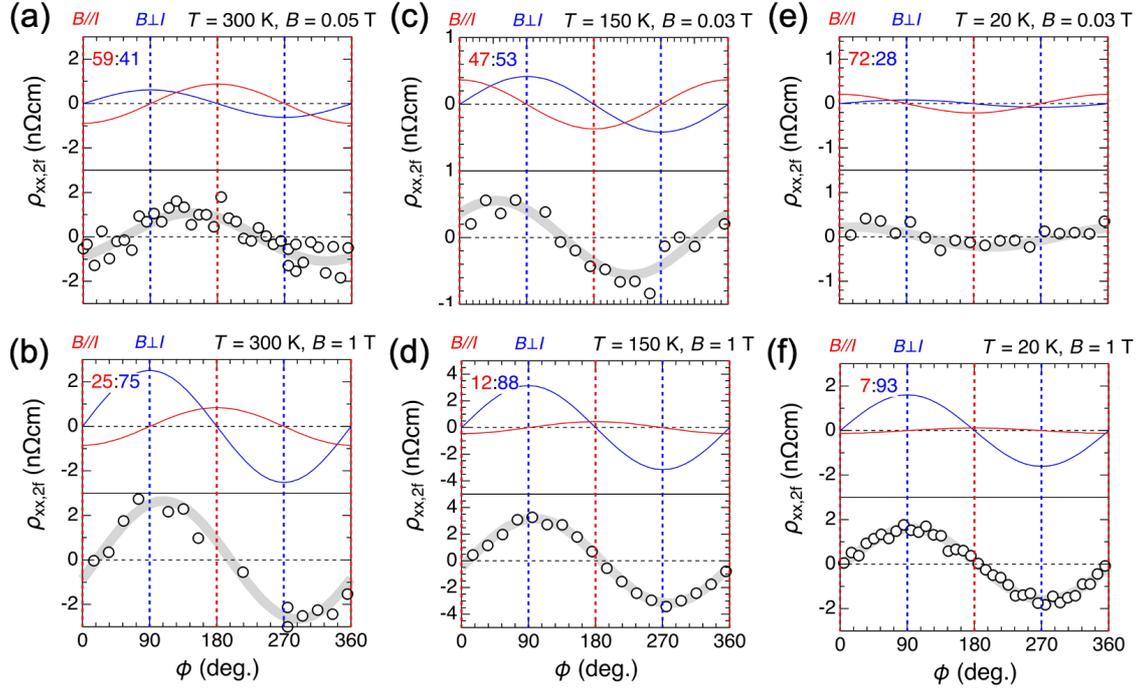

**Fig. S8:** Azimuth angle dependence of nonreciprocal resistivity at filled circles in Fig. S7(d). In the bottom panels, open symbols are the experimental data, which are fitted by the gray curves, $A_1\sin(\phi) + A_2\cos(\phi)$. In the top panels, decomposed curves, $A_1\sin(\phi)$ and $A_2\cos(\phi)$ that correspond to $\boldsymbol{I}\times\boldsymbol{B}$ and $\boldsymbol{I}\cdot\boldsymbol{B}$ terms, respectively, are separately plotted as blue and red curves. The ratios between $A_1$ and $A_2$ are also described at the left top part of each panel.

The $\boldsymbol{I}\cdot\boldsymbol{B}$ and $\boldsymbol{I}\times\boldsymbol{B}$ components in $\rho_{xx,2f}(\phi)$ are extracted as $\rho_{2f,I\cdot B}$ and $\rho_{2f,I\times B}$, and their temperature dependence is plotted in Figs. S9(a) and S9(b), respectively. The top, middle and bottom panels are the values in the conical, helical and forced ferromagnetic states shown in Fig. S7(d). We confirm that $\rho_{2f,I\cdot B}(T)$ is qualitatively similar to Figs. 3(c) and 3(d) in the main text, obtained for the microdevice #1. In Figs. 3(c) and 3(d) of the main text, the magnitude of $\rho^{2f}_{scat}$ significantly enhances towards $T_c$ = 300 K and exhibits a shoulder-like behavior around 100 K. At the transition field $B_{H\text{-}C}$, $\rho^{2f}_{band}$ takes a broad maximum around 150 K, and is suppressed around $T_c$. These behaviors are reproduced in Fig. S9(a), as expected. For $\rho_{2f,I\times B}$, the temperature dependence is qualitatively different from that of $\rho_{2f,I\cdot B}$. This component may arise from the different mechanism, such as observed in a polar semiconductor with the Rashba-type s-o coupling [8], either in the bulk or at the interface with the electrode.

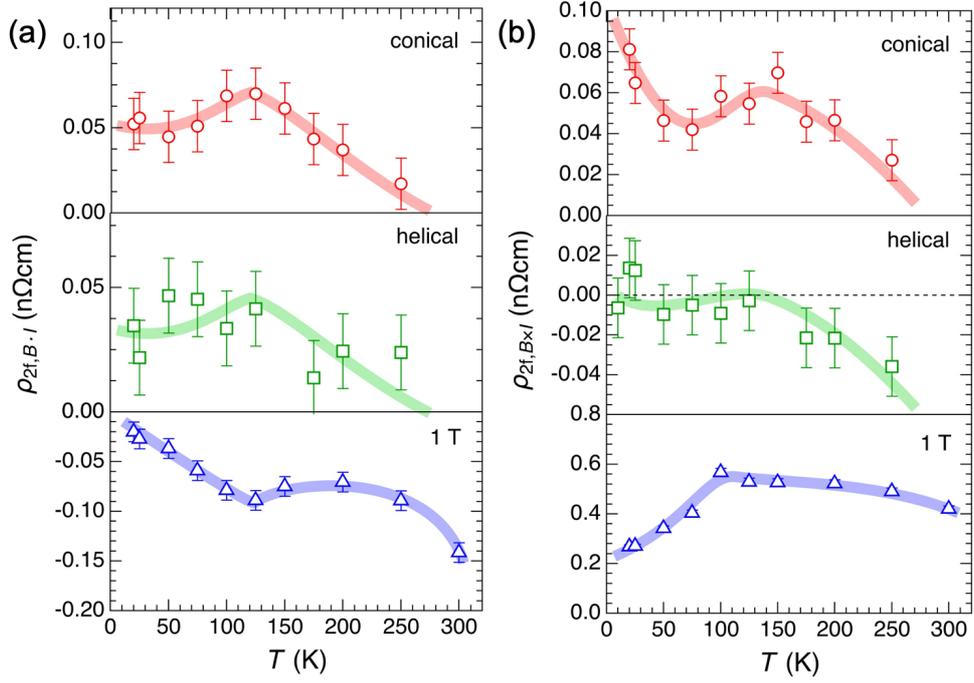

**Fig. S9:** Temperature dependence of the decomposed terms in nonreciprocal resistivity, $\rho_{2f,I\times B}(\phi = 90 \text{ deg.}) = A_1$ and $\rho_{2f,I\cdot B}(\phi = 0 \text{ deg.}) = A_2$. In the top, middle, and bottom panels, the results in the conical and helical phases, and 1 T shown in Fig. S7(d) are presented, respectively. The bold curves are the guide to the eye.

Next, to see whether the observed component $\rho_{2f,I\times B}$ is intrinsic to the bulk material or not, we rotate the magnetic field direction in the *yz*-plane of Fig. S7(c) (i.e. while keeping $\phi = 90$ deg.) and the polar angle ($\theta$) dependence of $\rho_{xx,2f}$ is measured at several values of temperature and magnetic field. As shown in Fig. S10, the experimental data for all the temperatures and fields can be fitted by $A\sin(\theta+\theta_0)$, clearly showing 360 deg. periodicity. According to the cubic crystal symmetry in $Co_8Zn_9Mn_3$, the four-fold rotational symmetry in $\theta$ should have been observed if the observed component $\rho_{2f,I\times B}$ were intrinsic bulk characteristics, due to the equivalence between the [010] and [001] axes. Therefore, we conclude that the $\rho_{2f,I\times B}$ observed in the $\phi$-dependence measurement is not an intrinsic property of the bulk. The voltage electrodes in the measurement sample were deposited on both the top ([001]) and side ([010]) surfaces, the former of which might contribute to the nonreciprocal resistivity stemming from Rashba-type s-o coupling near the interface without inversion symmetry.

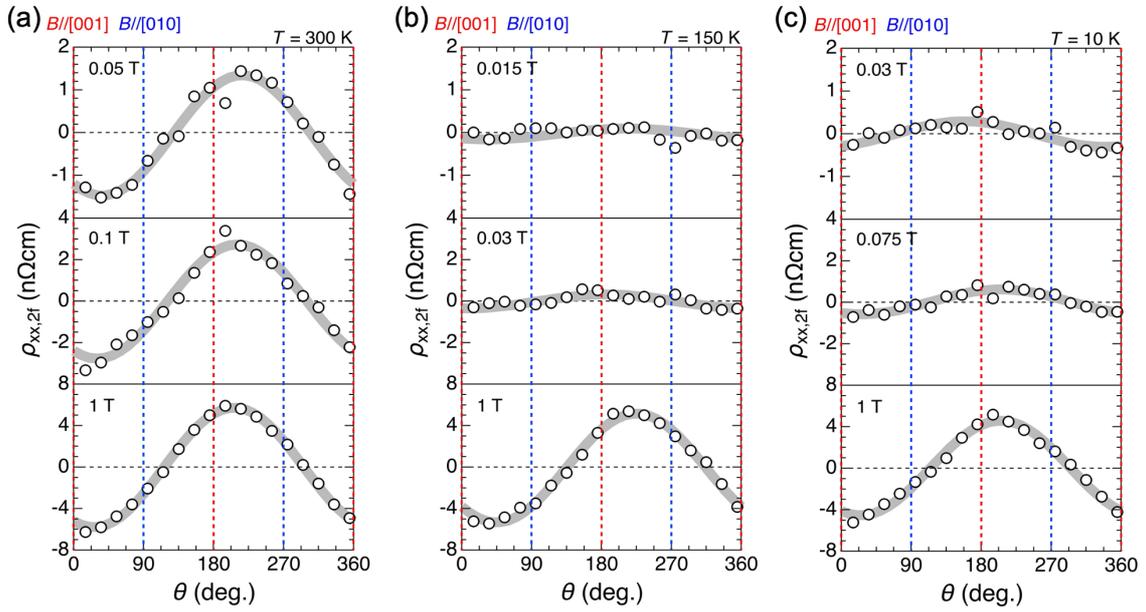

**Fig. S10:** Polar angle dependence of nonreciprocal resistivity at selected measurement conditions. Open symbols are the experimental data, which are fitted by the gray curves, $A\sin(\theta+\theta_0)$.

## D-4: Nonreciprocal resistivity – Theoretical calculation of the second harmonic resistivity

### D-4.1 Hamiltonian and eigenenergy

In this section, we explain a model to study the nonreciprocal second harmonic resistivity in the conical state of chiral magnets. We consider the Kondo-lattice model whose Hamiltonian is given by

$$H = \sum_{k} \varepsilon_k c_k^\dagger c_k - \sum_i c_i^\dagger (J\boldsymbol{S}_i - \boldsymbol{B}) \cdot \boldsymbol{\sigma} c_i, \tag{D1}$$

with the symbols defined in the main text. Note that $c_i = (c_{i\uparrow}, c_{i\downarrow})^T$ and we define the Fourier transform as $c_{i\sigma} = \frac{1}{\sqrt{N}}\sum_k e^{i\boldsymbol{k}\cdot\boldsymbol{r}_i} c_{k\sigma}$ with $N$ being the number of sites in the system. To calculate the eigenenergies, we first apply a unitary transformation, $c_k = U d_k$, with $U = \frac{1}{\sqrt{2}}\begin{pmatrix} 1 & -1 \\ 1 & 1 \end{pmatrix}$. After using the conical state of $\boldsymbol{S}_i$ defined in the main text, and introducing the factor $\zeta = \pm 1$ for magnetic field applied along the $\pm \hat{x}$ directions, the Hamiltonian becomes

$$H = \sum_k \varepsilon_k d_k^\dagger d_k - \sum_k (J\cos\alpha - B) d_k^\dagger \sigma_z d_k - \sum_k J\sin\alpha \left(d_{k\uparrow}^\dagger d_{k+\eta q\downarrow} + \text{c.c.}\right). \tag{D2}$$

This form clearly shows that the conical state with a helix wavevector $\boldsymbol{q}$ couples opposite-spin electrons of momentum difference $\eta\boldsymbol{q}$. Now we define another unitary transformation as $f_{k\uparrow} = d_{k-\frac{\eta q}{2}\uparrow}$, $f_{k\downarrow} = d_{k+\frac{\eta q}{2}\downarrow}$, and $f_k = (f_{k\uparrow}, f_{k\downarrow})^T$, which are local phase shifts as $f_{i\uparrow} = d_{i\uparrow} e^{i\eta\boldsymbol{q}\cdot\boldsymbol{r}_i/2}$ and $f_{i\downarrow} = d_{i\downarrow} e^{-i\eta\boldsymbol{q}\cdot\boldsymbol{r}_i/2}$. The Hamiltonian becomes

$$H = \sum_k f_k^\dagger \begin{pmatrix} \varepsilon_{k-\frac{\eta q}{2}} - \zeta(J\cos\alpha - B) & -J\sin\alpha \\ -J\sin\alpha & \varepsilon_{k+\frac{\eta q}{2}} + \zeta(J\cos\alpha - B) \end{pmatrix} f_k$$

$$\equiv \sum_k f_k^\dagger \begin{pmatrix} \varepsilon_-^{(\zeta)}(\boldsymbol{k}) & -J\sin\alpha \\ -J\sin\alpha & \varepsilon_+^{(\zeta)}(\boldsymbol{k}) \end{pmatrix} f_k. \tag{D3}$$

The eigenenergies are obtained as,

$$\mathcal{E}_\pm^{(\zeta)}(\boldsymbol{k}) = \frac{\varepsilon_+^{(\zeta)} + \varepsilon_-^{(\zeta)}}{2} \pm \sqrt{\left(\frac{\varepsilon_+^{(\zeta)} - \varepsilon_-^{(\zeta)}}{2}\right)^2 + (J\sin\alpha)^2}$$

$$= \frac{k^2 + q^2/4}{2m} \pm \sqrt{\left[\frac{\eta k_x q}{2m} + \zeta(J\cos\alpha - B)\right]^2 + J^2\sin^2\alpha}, \tag{D4}$$

as shown in the main text. For large $|\mathbf{k}|$, the eigenenergies become $\mathcal{E}_{\pm}^{(\zeta)}(\mathbf{k}) \xrightarrow{\text{large } |\mathbf{k}|} \frac{(k \pm \eta q/2)^2}{2m} \pm \zeta(J\cos\alpha - B) = \varepsilon_{\pm}^{(\zeta)}(\mathbf{k})$, while at $\mathbf{k} = 0$, the gap between two eigenenergies is $\mathcal{E}_{+}^{(\zeta)}(\mathbf{k}=0) - \mathcal{E}_{-}^{(\zeta)}(\mathbf{k}=0) = 2\sqrt{(J\cos\alpha - B)^2 + J^2 \sin^2\alpha} = 2J + \mathcal{O}(B/J)$. Thus the Kondo coupling introduces a hybridization of $\varepsilon_{\pm}^{(\zeta)}(\mathbf{k})$, as shown in Fig. 4(b) in the main text.

### D-4.2 Boltzmann transport theory for second harmonic resistivity

Following the approach in Ref. [8,27], we derive an analytical form for the second harmonic resistivity originating from the electronic band structure at zero temperature in this section. The Boltzmann equation is written as

$$-e\mathbf{E} \cdot \frac{\partial f}{\partial \mathbf{k}} = \frac{-(f - f_0)}{\tau}, \qquad \text{(D5)}$$

with $f$ and $f_0$ being the Fermi-Dirac distribution perturbed and unperturbed by external electric field $\mathbf{E} = E_x \hat{\mathbf{x}}$, respectively, and $\tau$ is an assumed single relaxation time. Defining a power expansion $f = \sum_{n=0,1,2,\ldots} f_n$ with $f_n \propto E_x^n$, we obtain $f_n = \left(e\tau E_x \frac{\partial}{\partial k_x}\right)^n f_0$ by comparing terms with the same order of $E_x$ on both sides of the Boltzmann equation. Following the similar procedure, we expand the current density up to the second order of $E_x$ as $J_x \approx J_{x,1} + J_{x,2} \equiv \sigma_1 E_x + \sigma_2 E_x^2$, with the form (for brevity, we ignore the field-direction superscript $(\zeta)$ in $\mathcal{E}_\pm$ below)

$$J_{x,1} = \sum_\pm \int \frac{d^3k}{(2\pi)^3}\left(-e\frac{\partial \mathcal{E}_\pm}{\partial k_x}\right) f_{1,\pm} = e^2 \tau E_x \sum_\pm \int \frac{d^3k}{(2\pi)^3} \frac{\partial^2 \mathcal{E}_\pm}{\partial k_x^2} f_{0,\pm} \equiv \sigma_1 E_x, \quad \text{(D6)}$$

$$J_{x,2} = \sum_\pm \int \frac{d^3k}{(2\pi)^3}\left(-e\frac{\partial \mathcal{E}_\pm}{\partial k_x}\right) f_{2,\pm} = -e^3 \tau^2 E_x^2 \sum_\pm \int \frac{d^3k}{(2\pi)^3} \frac{\partial^3 \mathcal{E}_\pm}{\partial k_x^3} f_{0,\pm} \equiv \sigma_2 E_x^2, \quad \text{(D7)}$$

where in the second equalities of both lines we have used integration by parts, $f_{0,\pm} = 1/(\exp[\beta(\mathcal{E}_\pm - \mu)] + 1)$ with $\beta = 1/k_B T$ ($k_B$ is the Boltzmann constant and $T$ is the temperature), $\mu$ is the chemical potential, and we have summed over the contributions from the two bands denoted by $\sum_\pm (\ldots)$.

Assuming the second harmonic resistivity contains only $\mathbf{J} \cdot \mathbf{B}$ term based on Onsager symmetry argument [1], we write

$$E_x = \rho_{xx,1f} J_x + \rho_{xx,1f} \gamma (\mathbf{J} \cdot \mathbf{B}) J_x/2 = \rho_{xx,1f} \sigma_1 E_x + \rho_{xx,1f}(\sigma_2 + \gamma B \sigma_1^2/2) E_x^2 + \mathcal{O}(E_x^3),$$

$$\Rightarrow \rho_{xx,1f} = \frac{1}{\sigma_1}, \quad \gamma = \frac{-2\sigma_2}{B\sigma_1^2}, \quad (D8)$$

which is valid up to second order of $E_x$. In our experiment, the second harmonic voltage $V_{2f}$ is described by $V = \rho_{xx,1f} \gamma B j_0^2 \sin^2(\omega t) d/2 = V_0 + V_{2f} \cos(2\omega t)$ with $d$ being the length between voltage terminals. The measured second harmonic resistivity is defined by

$$\rho^{2f} = \frac{V_{2f}}{j_0 d} = \frac{-\rho_{xx,1f} \gamma B j_0}{4} = \frac{\sigma_2}{2\sigma_1^3} j_0. \quad (D9)$$

From Eqs. (D6–D7), the zero-temperature first- and second-order currents are proportional to the $\mathbf{k}$-space integral of the second- and third-order derivatives of $\mathcal{E}_\pm^{(\zeta)}$ by $k_x$, respectively. Here we consider the origin of the nonreciprocal $\rho_{\text{band}}^{2f}$. Due to a symmetry property of eigenenergy Eq. (D4) between opposite field directions, $\mathcal{E}_\pm^{(-)}(k_x) = \mathcal{E}_\pm^{(+)}(-k_x)$ with fixed $k_y$ and $k_z$, we get $d^2\mathcal{E}_\pm^{(-)}(k_x)/dk_x^2 = d^2\mathcal{E}_\pm^{(+)}(-k_x)/d(-k_x)^2$ and $d^3\mathcal{E}_\pm^{(-)}(k_x)/dk_x^3 = -d^3\mathcal{E}_\pm^{(+)}(-k_x)/d(-k_x)^3$. Therefore, the integrands of first (second)-order currents between opposite fields are related by a positive (negative) mirror symmetric transformation relative to $k_x$. From Eq. (D4), we can also derive $d^3\mathcal{E}_+^{(\zeta)}(k_x)/dk_x^3 = -d^3\mathcal{E}_-^{(\zeta)}(k_x)/dk_x^3$, thus the two bands have contributions opposite in sign. However, the Fermi wavevectors of the two bands are different, as denoted by vertical green and red dotted lines in Fig. S11(e–h), such that there is no cancellation of contributions by the two bands. Therefore, as shown by comparing Fig. S11(e) with (f) and (g) with (h), the part of net second-order currents for opposite fields from an integration path of $k_y = k_z = 0$ and with $k_x$ within the range inside vertical dotted lines will be finite and opposite in sign for opposite magnetic fields. This is our proposed possible mechanism for the nonreciprocity stemming from band asymmetry, which in our case is due to the Kondo coupling of electron spins with the conical state magnetizations.

### D-4.3 Calculation of $\rho_{\text{band}}^{2f}$

The calculations of $J_{x,1}$ and $J_{x,2}$ in Eq. (D6) and (D7) require the second- and third-order derivatives of eigenenergy, respectively, which have the form

$$\frac{\partial^2 \varepsilon_\pm}{\partial k_x^2} = \frac{1}{m} \pm \frac{q^2 (J \sin \alpha)^2}{4m^2 \left\{ \left[\frac{\eta k_x q}{2m} + \zeta(J \cos \alpha - B)\right]^2 + (J \sin \alpha)^2 \right\}^{3/2}}, \quad \text{(D10)}$$

$$\frac{\partial^3 \varepsilon_\pm}{\partial k_x^3} = \mp \frac{3\eta q^3 (J \sin \alpha)^2 \left[\frac{\eta k_x q}{2m} + \zeta(J \cos \alpha - B)\right]}{8m^3 \left\{ \left[\frac{\eta k_x q}{2m} + \zeta(J \cos \alpha - B)\right]^2 + (J \sin \alpha)^2 \right\}^{5/2}}. \quad \text{(D11)}$$

Considering the small value of $q$ (about $2\pi/10^2$ nm$^{-1}$) compared with typical Fermi wave numbers (for Fermi energy around few eV with bare electron mass) and the small value of $B$ compared with $J \cos \alpha$ as shown in the inset of Fig. S12, we calculate the zero-temperature $\rho_{\text{band}}^{2f}$ up to the lowest order of $q$ and ignore $B$ in the factor $(J \cos \alpha - B)$. Because of the form of Eq. (D9), we only keep the first term $1/m$ with zeroth order in $q$ in Eq. (D10) to calculate $\sigma_1$ via Eq. (D6). Also, the zero-temperature Fermi-Dirac distribution which become step functions as $f_{0,\pm}(k) = \Theta\left(k_{F,\pm} - k\right)$ in integrands of Eqs. (D6)–(D7) are also expanded in orders of $q$ and only the zeroth-order terms are retained, namely, $\varepsilon_\pm\left(k_{F,\pm}\right) \approx \frac{k_{F,\pm}^2}{2m} \pm J = \mu$ and thus $k_{F,\pm} = \sqrt{2m(\mu \mp J)}$. The integrals reduce to

$$J_{x,1} \approx \frac{e^2 \tau E_x}{m} \sum_\pm \int \frac{d^3 k}{(2\pi)^3} f_{0,\pm} = \frac{e^2 \tau E_x}{6\pi^2 m} \left(k_{F,+}^3 + k_{F,-}^3\right) = \sigma_1 E_x, \quad \text{(D12)}$$

$$J_{x,2} \approx \frac{3\zeta \eta e^3 \tau^2 E_x^2 q^3 \cos \alpha \sin^2 \alpha}{8 m^3 J^2} \int \frac{d^3 k}{(2\pi)^3} (f_{0,+} - f_{0,-})$$

$$= \frac{\zeta \eta e^3 \tau^2 E_x^2 q^3 \cos \alpha \sin^2 \alpha}{16 \pi^2 m^3 J^2} \left(k_{F,+}^3 - k_{F,-}^3\right) = \sigma_2 E_x^2. \quad \text{(D13)}$$

It is apparent that $J_{x,2}$ changes sign when magnetic field direction is reversed to cause sign change of $\zeta$. This causes the finite nonreciprocal $\rho_{\text{band}}^{2f}$. From Eqs. (D8) and (D9) we get the analytical expressions

$$\rho_{xx,1f} = \frac{1}{\sigma_1} = \frac{6\pi^2 m}{\tau e^2 \left(k_{F,+}^3 + k_{F,-}^3\right)}, \quad \text{(D14)}$$

$$\rho_{\text{band}}^{2\text{f}} = \frac{27\pi^4 \zeta \eta j_0 q^3 (k_{F,+}^3 - k_{F,-}^3)}{4\tau e^3 J^2 (k_{F,+}^3 + k_{F,-}^3)^3} \cos\alpha(B) \sin^2\alpha(B), \quad (D15)$$

$$\gamma = -\frac{9\pi^2 \zeta \eta q^3 (k_{F,+}^3 - k_{F,-}^3)}{2meBJ^2 (k_{F,+}^3 + k_{F,-}^3)^2} \cos\alpha(B) \sin^2\alpha(B). \quad (D16)$$

The proportionality of $\cos\alpha(B)\sin^2\alpha(B)$ is used in Fig. 4(c) of main text to compare the low-temperature experiment.

Here we note an intriguing property of the $\rho_{\text{band}}^{2\text{f}}$ in Eq. (D15). In [13], the Kondo coupling between conduction electron spin and local magnetizations is taken as a perturbation used to calculate the asymmetric scattering channel in the first Born approximation of Boltzmann equation, and the resulting $\rho_{2\text{f}}$ is proportional to $M\langle\chi_x\rangle$ when the electric field is applied in the $x$ direction. The formula we find in Eq. (D15) which comes from an exact diagonalization of the Hamiltonian containing the Kondo coupling with conical state, is actually also proportional to $M\langle\chi_x^{\text{conical}}\rangle$ with $\chi_x^{\text{conical}}$ being the vector chirality induced by the conical state. To see this, first we note that $\cos\alpha(B) \propto M(B)$ by our definition. We then consider a conical state in the discrete space as

$$\boldsymbol{S}_n = \sin\alpha [\hat{\boldsymbol{y}}\cos(qx_n) + \eta\hat{\boldsymbol{z}}\sin(qx_n)] + \cos\alpha\hat{\boldsymbol{x}}, \quad x_n = na_0, \quad (D17)$$

with $a_0$ being the lattice constant and $n$ being integers. Supposed there are $N_x$ sites in the $x$ direction, the averaged vector chirality is calculated as

$$\langle\chi_x^{\text{conical}}\rangle = \frac{1}{N_x}\sum_{n=1}^{N_x}(\boldsymbol{S}_n \times \boldsymbol{S}_{n+1})_x$$

$$= \frac{1}{N_x}\eta\sin^2\alpha\sum_{n=1}^{N_x}[\cos(qna_0)\sin(q(n+1)a_0) - \sin(qna_0)\cos(q(n+1)a_0)]$$

$$= \eta\sin(qa_0)\sin^2\alpha. \quad (D18)$$

Therefore, $\sin^2\alpha \propto \langle\chi_x^{\text{conical}}\rangle$, and we get from Eq. (D15) the relation $\rho_{\text{band}}^{2\text{f}} \propto \cos\alpha\sin^2\alpha \propto M\langle\chi_x^{\text{conical}}\rangle$, similar to the relation of only up to the first order of Born approximation.

**D-4.4 Calculation of magnon-induced vector chirality**

In this section, we calculate the vector chirality fluctuation induced by magnon excitations at low temperatures in the forced ferromagnetic state under high magnetic fields, in order to compare with $\rho_{\text{scat}}^{2\text{f}}$ at high fields below $T_c$ measured in our experiment. Consider the Hamiltonian for local magnetizations $\boldsymbol{S}_i$ located at site $i$, ignoring the interaction with itinerant electron spins for simplicity,

$$H = -J_F \sum_{\langle ij \rangle} \boldsymbol{S}_i \cdot \boldsymbol{S}_j - \boldsymbol{B} \cdot \sum_i \boldsymbol{S}_i + D \sum_i \hat{\boldsymbol{z}} \cdot \boldsymbol{S}_i \times \boldsymbol{S}_{i+z}, \quad (D19)$$

where the three terms on the right-hand side are ferromagnetic nearest-neighbor exchange interaction, Zeeman energy, and DMI, respectively. For convenience regarding the common form of magnon Holstein-Primakoff transformation, we consider the magnetic field as applied in the $\hat{\boldsymbol{z}}$ direction, $\boldsymbol{B} = B\hat{\boldsymbol{z}}$, and assume the helical vector $\boldsymbol{q} = q\hat{\boldsymbol{z}}$, such that only the $\hat{\boldsymbol{z}}$ component of the DMI is considered. The following equations apply for our experimental situation after changing coordinates $z \to x, x \to y,$ and $y \to z$. For the field applied in the $+\hat{\boldsymbol{z}}$ direction, the Holstein-Primakoff transformations up to the second-order magnon expansion are $S_i^z = S - a_i^\dagger a_i$, $S_i^+ = S_i^x + iS_i^y \approx \sqrt{2S} a_i$, and $S_i^- = S_i^x - iS_i^y \approx \sqrt{2S} a_i^\dagger$, with $a_i$ being the magnon annihilation operator at site $i$ satisfying Boson commutation relation $[a_i, a_j^\dagger] = \delta_{i,j}$. Using the Fourier transform, $a_{\boldsymbol{k}} = \frac{1}{\sqrt{N}} \sum_i e^{-i\boldsymbol{k} \cdot \boldsymbol{r}_i} a_i$ with $\boldsymbol{r}_i$ being the position of site $i$ and $N$ the number of sites in the system, the Hamiltonian can be written as [S2],

$$H = \sum_{\boldsymbol{k}} \omega_{\boldsymbol{k}} a_{\boldsymbol{k}}^\dagger a_{\boldsymbol{k}},$$
$$\omega_{\boldsymbol{k}} = 2J_F S[3 - \sum_{j=x,y,z} \cos(k_j a_0)] + B - 2SD\,\text{sgn}(\boldsymbol{k} \cdot \boldsymbol{B})|\sin(k_z a_0)|, \quad (D20)$$

where $a_0$ is the lattice constant, and in the last DMI contribution in $\omega_{\boldsymbol{k}}$ the dependence on the sign of $\boldsymbol{k} \cdot \boldsymbol{B}$ comes from the different transformation $S_i^z = -S + a_i^\dagger a_i$, $S_i^+ = \sqrt{2S} a_i^\dagger$, $S_i^- = \sqrt{2S} a_i$ when magnetic field is reversed to the $-\hat{\boldsymbol{z}}$ direction. In the following we consider the case with magnetic field applied in the $+\hat{\boldsymbol{z}}$ direction. The averaged vector chirality can be written in terms of the magnon number as

$$\chi_z = \frac{1}{N} \sum_i (\boldsymbol{S}_i \times \boldsymbol{S}_{i+z})_z = \frac{-2S}{N} \sum_{\boldsymbol{k}} \sin(k_z a_0)\, a_{\boldsymbol{k}}^\dagger a_{\boldsymbol{k}}, \quad (D21)$$

which is simply the DMI with $D$ replaced by $1/N$. To calculate the fluctuation of $\chi_z$, we use the thermal average of magnon number as $\langle a_{\boldsymbol{k}}^\dagger a_{\boldsymbol{k}} \rangle = 1/(e^{\beta \omega_{\boldsymbol{k}}} - 1)$. For Co-Zn-Mn alloys, the ratio of $D$ to $J_F$ is roughly in the order of $10^{-2}$ [S3–S5], thus we expand the $\omega_k$ up to first order of $D$ and use the continuous $\boldsymbol{k}$-integral,

$$\langle \chi_z \rangle \approx -2Sa_0^3 \int \frac{d^3k}{(2\pi)^3} \sin(k_z a_0) \left[ \frac{1}{e^{\beta\omega_0}-1} + \left(\frac{d}{d\omega_0}\frac{1}{e^{\beta\omega_0}-1}\right)[-2SD\sin(k_z a_0)] \right]$$

$$= 4S^2 D a_0^3 \int \frac{d^3k}{(2\pi)^3} \sin^2(k_z a_0) \left(\frac{d}{d\omega_0}\frac{1}{e^{\beta\omega_0}-1}\right), \quad (D22)$$

where $\omega_0(\boldsymbol{k}) = 2J_F S[3 - \sum_j \cos(k_j a_0)] + B$ is even in $\boldsymbol{k}$, such that the first integral in bracket in the first line is zero due to the oddness of $\sin(k_z a_0)$ in $k_z$. We thus observe that without DMI (i.e. without s-o interaction) the magnons cannot induce finite vector chirality fluctuations. In the continuous limit $a_0 \to 0$, we set $\omega_0 \to J_F S a_0^2 k^2$ ignoring the magnetic field since $\mu_B B/J_F S \sim 10^{-4}$ even at a high field of $B = 0.1$ T, and set $\sin(k_z a_0) \to k_z a_0$. The averaged vector chirality becomes

$$\langle \chi_z \rangle \approx \frac{-4S^2 D a_0^5 \beta}{(2\pi)^3} \int d^3k \, k_z^2 \frac{e^{\beta J_F S a_0^2 k^2}}{\left(e^{\beta J_F S a_0^2 k^2}-1\right)^2} = \frac{-4S^2 D a_0^5 \beta}{(2\pi)^3} \frac{4\pi}{3} \int dk \, k^4 \frac{e^{\beta J_F S a_0^2 k^2}}{\left(e^{\beta J_F S a_0^2 k^2}-1\right)^2} \approx$$

$$\frac{-S^2 D a_0^5}{3\pi^2} \frac{\beta}{\alpha^{5/2}} \int_0^\infty dx \frac{x^{3/2} e^x}{(e^x-1)^2} = \frac{-3.47 D (k_B T)^{3/2}}{3\pi^2 J_F^{5/2} S^{1/2}}, \quad (D23)$$

where in the third equality we have used $x \equiv \alpha k^2, \alpha \equiv \beta J_F S a_0^2$, and the approximation is done by setting the maximum of $x$ as infinity in the low temperature limit, with the numerical integral value of about 3.47. This low-temperature, continuous limit of the vector chirality has a temperature dependence of $T^{3/2}$, and it depends on the sign of DMI constant $D$. We also calculate numerically the vector chirality up to first-order of DMI including the Zeeman term in $\omega_0$ without taking continuous and low-temperature limit. The analytical and numerical values of $\langle \chi_z \rangle$ are plotted as blue curves in Fig. 4(d) to compare with the $\rho_{\text{scat}}^{2f}$ observed in our experiment below $T_c$. Since the analytical magnon chirality shows a $T^{3/2}$ dependence, the blue dotted curve is obtained by calculating the coefficient $c$ of a least square fit of the form $\rho_{\text{scat}}^{2f} = cT^{3/2}$ to fit the experimental data. The fitted coefficient $c$ is also multiplied by the numerical integral solution of Eq. (D21) divided by its approximate analytical form Eq. (D23) to obtain the blue solid curve. Figure 4(d) shows a qualitatively good agreement between experiment and theory, indicating that $\rho_{\text{scat}}^{2f}$ at $T < 280$ K may be attributed to the magnon-induced vector chirality.

Here we calculate the averaged magnon number as a function of temperature in the same manner up to the first order of DMI,

$$\frac{1}{N}\sum_{\mathbf{k}}\langle a_{\mathbf{k}}^{\dagger}a_{\mathbf{k}}\rangle \approx a_0^3 \int \frac{d^3k}{(2\pi)^3}\left[\frac{1}{e^{\beta\omega_0}-1} + \left(\frac{d}{d\omega_0}\frac{1}{e^{\beta\omega_0}-1}\right)[-2SD\sin(k_z a_0)]\right]. \quad (D24)$$

The integral of the second term due to DMI in bracket is zero since it is an odd function of $k_z$. We thus get

$$\frac{1}{N}\sum_{\mathbf{k}}\langle a_{\mathbf{k}}^{\dagger}a_{\mathbf{k}}\rangle \approx \frac{a_0^3}{2\pi^2}\int dk \frac{k^2}{e^{\beta J_F S a_0^2 k^2}-1} \approx \frac{a_0^3}{(2\pi)^2 \alpha^{3/2}}\int_0^{\infty} dx \frac{x^{1/2}}{e^x-1} = \frac{2.32}{(2\pi)^2}\left(\frac{k_B T}{J_F S}\right)^{3/2},$$
(D25)

which is proportional to $T^{3/2}$ as the usual ferromagnetic Heisenberg model since up to the first order of DMI it has no contribution. This magnon number contributes to the approximate $T^{3/2}$ dependence of $\rho_{\text{scat}}^{2\text{f}}$ at $T < 280$ K, which is also plotted versus temperature in Fig. S13 for reference.

**SI References:**

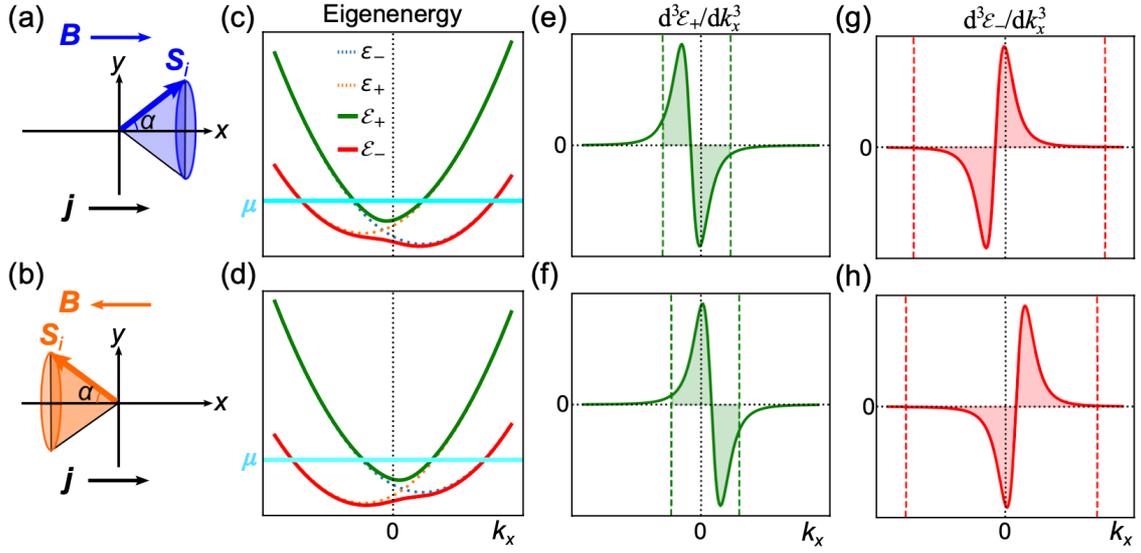

**Fig. S11:** Schematics of (a,b) the conical spin states (c,d) energy band dispersions shown in Figs. 4(a) and 4(b) in the main text for reference, and (e–h) third-order derivatives of the energy dispersions. Figures (c), (e), and (g) are for those under application of a magnetic field along the $+\hat{x}$ direction, while (d), (f), and (h) are for the field along the $-\hat{x}$ direction. In (e–h), the vertical green and red dotted lines indicate the points of $k_x$ at which the chemical potential μ intersects the energy bands in (c,d), and the shaded areas denote the contribution of second-order current in the integration path with $k_y = k_z = 0$ and $k_x$ within the range inside vertical colored dotted lines.

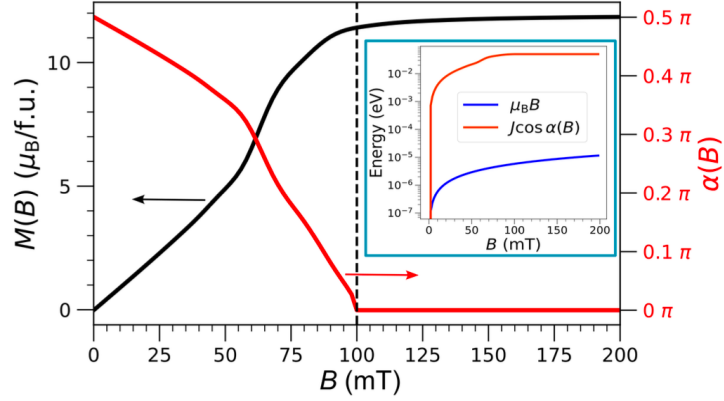

**Fig. S12:** Experimental magnetization curve $M(B)$ at 10 K (black curve) and the magnetization tilt angle $\alpha(B)$ (red curve) along the magnetic field direction extracted from defining $\cos\alpha(B) = M(B)/M(0.1\text{ T})$ by assuming the saturation of $M$ as at $B = 0.1$ T (vertical dotted line). The inset shows a comparison of the magnitudes of $\mu_B B$ (blue curve) and $J\cos\alpha(B)$ (red curve) taking a typical value of $J = 0.05$ eV.

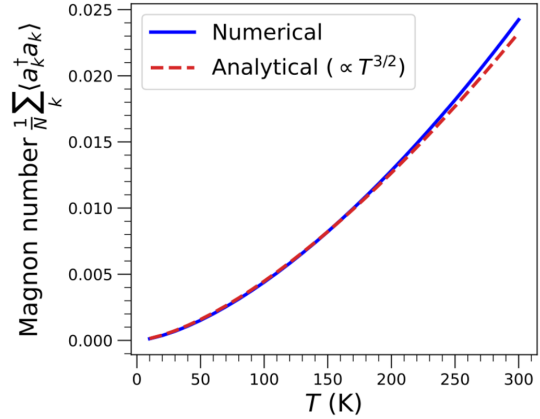

**Fig. S13:** Calculated analytical (red dotted) and numerical (blue curve) magnon numbers from Eq. (D24).